%% file: main.tex
\documentclass[sigconf]{acmart}

\AtBeginDocument{%
  \providecommand\BibTeX{{%
    \normalfont B\kern-0.5em{\scshape i\kern-0.25em b}\kern-0.8em\TeX}}}

\usepackage{float}
\usepackage{subfigure}
\usepackage{bbding}
\usepackage{fancyvrb}
\usepackage{graphicx}
\usepackage{geometry}
\usepackage{makecell}
\usepackage{color}
\usepackage{multirow}
\usepackage{subcaption}
\usepackage{soul}
\usepackage{stfloats}
\usepackage{enumitem}

\definecolor{lightred}{HTML}{D89090}
\definecolor{darkblue}{HTML}{104E8B}
\definecolor{midblue}{HTML}{376B9E}
\definecolor{lightblue}{HTML}{5F89B1}
\definecolor{newgreen}{HTML}{C5E9E3}

\newif\iffinalversion
\finalversiontrue         

\iffinalversion
  \newcommand{\added}[1]{#1}       
  \newcommand{\deleted}[1]{}       
\else
  \newcommand{\added}[1]{\textcolor{blue}{#1}}
  \newcommand{\deleted}[1]{\textcolor{red}{\sout{#1}}}
\fi

\newcommand{\system}{Vistoria}

\begin{document}

\author{Kexue Fu}
\orcid{0000-0002-2929-2663}
\authornote{These authors contributed equally to this work.}
\authornote{This work was conducted while Kexue Fu was a visiting scholar at the University of Notre Dame.}

\affiliation{
  \institution{City University of Hong Kong}
  \department{School of Creative Media}
  \city{Hong Kong, SAR}
  \country{China}
}

\affiliation{
  \institution{University of Notre Dame}
  \department{Department of Computer Science and Engineering}
  \city{Notre Dame}
  \state{Indiana}
  \country{USA}
}
\email{kexuefu2-c@my.cityu.edu.hk}

\author{Jingfei Huang}
\authornotemark[1]
\orcid{0009-0002-0213-4160}
\affiliation{
  \institution{Harvard University}
    \department{Graduate School of Design \& School of Engineering and Applied Science}
  \city{Cambridge}
  \state{Massachusetts}
  \country{USA}
}
\email{jingfeihuang@mde.harvard.edu}

\author{Long Ling}
\authornotemark[1]
\orcid{0009-0001-2635-788X}
\affiliation{
  \institution{Tongji University}
  \department{College of Design and Innovation}
  \city{Shanghai}
  \country{China}
}
\email{lucyling0224@gmail.com}

\author{Sumin Hong}
\orcid{0000-0002-4671-7192}
\affiliation{
  \institution{University of Notre Dame}
  \department{Department of Computer Science and Engineering}
  \city{Notre Dame}
  \state{Indiana}
  \country{USA}
}
\email{shong6@nd.edu}

\author{Yihang Zuo}
\orcid{0000-0002-6843-8194}
\affiliation{
  \institution{The Hong Kong University of Science and Technology (Guangzhou)}
  \city{Guangzhou}
  \state{Guangdong}
  \country{China}
}
\email{yzuo099@connect.hkust-gz.edu.cn}

\author{RAY LC}
\authornotemark[3]
\orcid{0000-0001-7310-8790}
\affiliation{
  \institution{City University of Hong Kong}
  \department{School of Creative Media}
  \city{Hong Kong, SAR}
  \country{China}
}
\email{ray.lc@cityu.edu.hk}

\author{Toby Jia-Jun Li}
\orcid{0000-0001-7902-7625}
\authornote{Correspondences can be addressed to toby.j.li@nd.edu and ray.lc@cityu.edu.hk.}
\affiliation{
  \institution{University of Notre Dame}
  \department{Department of Computer Science and Engineering}
  \city{Notre Dame}
  \state{Indiana}
  \country{USA}
}
\email{toby.j.li@nd.edu}


\copyrightyear{2026}
\acmYear{2026}
\setcopyright{cc}
\setcctype{by}
\acmConference[CHI '26]{Proceedings of the 2026 CHI Conference on Human Factors in Computing Systems}{April 13--17, 2026}{Barcelona, Spain}
\acmBooktitle{Proceedings of the 2026 CHI Conference on Human Factors in Computing Systems (CHI '26), April 13--17, 2026, Barcelona, Spain}
\acmDOI{10.1145/3772318.3790400}
\acmISBN{979-8-4007-2278-3/2026/04}

\title[\system]{\system: A Multimodal System to Support Fictional Story Writing through Instrumental Image-Text Co-Editing}

\begin{abstract}

Humans think visually---we remember in images, dream in pictures, and use visual metaphors to communicate. Yet, most creative writing tools remain text-centric, limiting how writers plan and translate ideas. We present Vistoria, a system for synchronized image-text co-editing in fictional story writing. A formative Wizard-of-Oz co-design study with 10 story writers revealed how sketches, images, and text serve as essential elements for ideation and organization. Drawing on theories of Instrumental Interaction, Vistoria introduces instrumental operations---Lasso, Collage, Perspective Shift, and Filter that enable seamless narrative exploration across modalities. A controlled study with 12 participants shows that co-editing enhances expressiveness, immersion, and collaboration, opening space for writers to follow divergent story directions and craft more vivid, detailed narratives. While multimodality increased cognitive demand, participants reported stronger senses of ownership and agency. These findings demonstrate how multimodal co-editing expands creative potential by balancing abstraction and concreteness in narrative development.
\end{abstract}




\begin{teaserfigure}
  \centering
  \includegraphics[width=0.85\columnwidth,keepaspectratio]{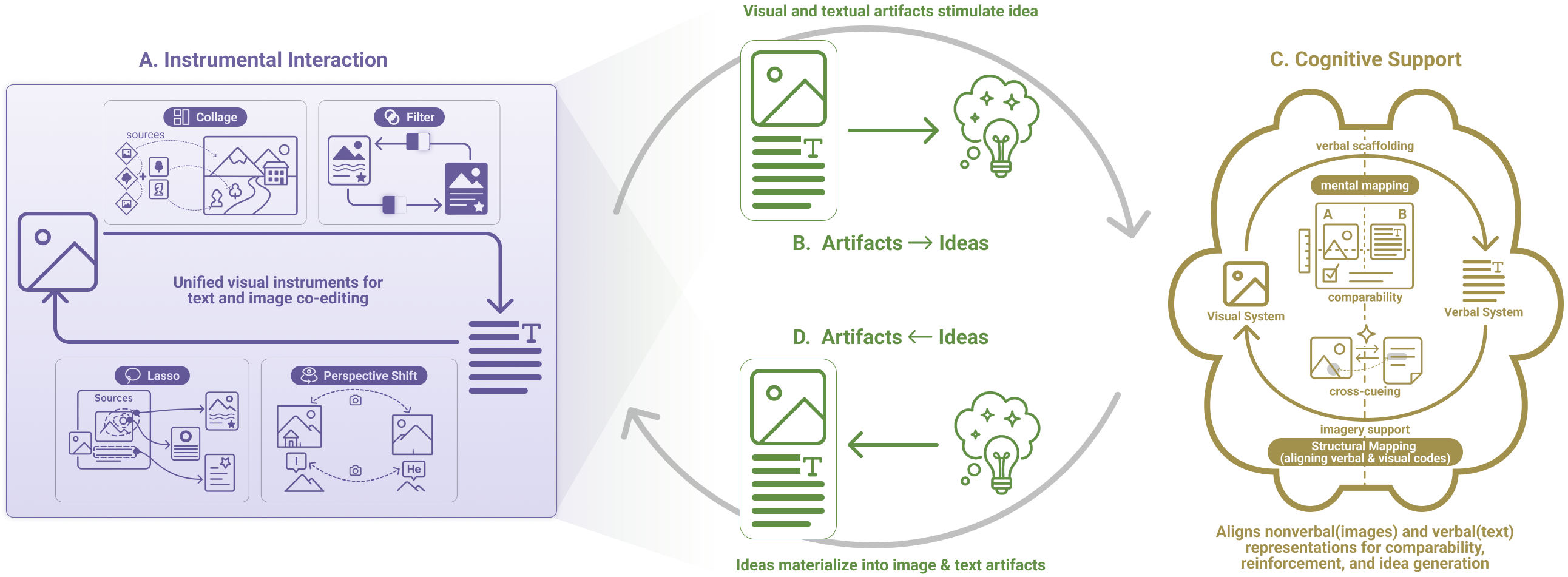}
  \caption{Vistoria supports a cyclic workflow in which multimodal artifacts and ideas co-evolve.
(A) Instrumental operations enable image–text co-editing.
(B) Resulting image–text artifacts prompt new story directions.
(C) Image–text alignment supports idea formation by coordinating verbal and non-verbal processing.
(D) Emerging ideas are externalized into new artifacts, closing the loop for iterative ideation and integration.}
  \label{fig: teaser}
  \Description{The figure illustrates Vistoria’s overall cyclic workflow for multimodal story writing. It shows how writers interact with image–text instruments to edit visual and textual artifacts, how these artifacts stimulate new story ideas, and how ideas are materialized back into new image–text cards. The cycle highlights continuous co-evolution between artifacts and ideas, emphasizing synchronized visual and verbal processing during iterative narrative development.}
\end{teaserfigure}

\maketitle

\section{Introduction}\label{sec:Introduction}

\input{sections/01-Intro}

\section{Related Work}\label{sec:Background}
\input{sections/02-Related-Work}

\section{Formative Study}\label{sec:Design and Development}
\input{sections/03-Formative-Study}

\section{\system{} System}\label{sec:Methods}

\input{sections/04-system}

\section{User Study}\label{sec:Results}

\input{sections/05-User-Study}

\section{Discussion}\label{sec:Discussion}
\input{sections/06-Discussion}

\section{Conclusion}\label{sec:Conclusion}
\input{sections/07-Conclusion}

\section*{\added{Acknowledgments}}
\added{This work is supported by City University of Hong Kong Teaching Development Grant (6000901), City University of Hong Kong Booster Fund (7030021), and Hong Kong Research Grants Council Theme-based Research Scheme (T45-205/21-N).
We sincerely thank all anonymous reviewers for their insightful comments and constructive feedback that helped improve this paper.} 

\newpage
\bibliographystyle{ACM-Reference-Format}
\balance
\bibliography{references}
\clearpage
\onecolumn 
\appendix
\section{Appendix}
\label{sec:Appendix}
\input{sections/Appendix}

\end{document}

%% file: sections/01-Intro.tex
Human thinking involves multimodal processing. Visual processes play a central role in cognition: we recall experiences as spatial scenes, form mental models through imagery, and use visual structure to organize and interpret information\added{
~\cite{rezaee2011investigating, huang2024sight, kanellopoulou2019dual}}. Language is similarly entwined with imagery. Particularly, text comprehension often evokes mental pictures, and abstract ideas are commonly articulated through spatial metaphors such as path, framework, or perspective~\cite{zhaoMakingWriteConnections2025,clark1991dual,gardenfors1996mental}. Dual Coding Theory frames this coupling between imagery and language, positing that humans draw on both verbal and nonverbal channels to represent and support specific reasoning and communicative processes~\cite{clark1991dual}.


Story writing is particularly multimodal in nature. During the planning phase, experienced writers often use both imagery and language to construct the story world. They visualize spatial layouts, character interactions, and scene dynamics, while using textual notes to label, sequence, and reason about narrative structure~\cite{doyle1998writer,bal2013does,chungToytellerAIpoweredVisual2025a,dangWorldSmithIterativeExpressive2023}. In the translating phase, visual details serve as an anchor that shapes how writers organize narrative detail and emotional tone, while texts linearize these visualized ideas into descriptions, dialogue, and narrative perspective that readers can follow~\cite{mukramah2023effect,listyani2019use}. Crucially, nonverbal and verbal channels do not operate in isolation: writers use imagery to trigger new wording, and emerging text in turn elicits further mental images~\cite{zhaoMakingWriteConnections2025,lee2024design}. Therefore, there is a great opportunity for tools that support story writing to match this multimodal complexity by accommodating the continuous interplay between visual and textual thinking within the same workflow.
Yet, current writing tools remain overwhelmingly text-centric, treating linear text as the primary or sole medium of expression~\cite{lee2024design,geroSocialDynamicsAI2023}. Although recent systems powered by Large Language Models (LLMs) incorporate visual elements through image generation or retrieval, these visuals remain peripheral. They function mainly as prompts~\cite{qin2024charactermeet}, static references~\cite{raoScriptVizVisualizationTool2024}, or organizational diagrams~\cite{mishraWhatIFBranchedNarrative2025,ClueCartSupportingGame}, rather than as tightly integrated, manipulable representations along with text. This requires writers to translate visual ideas back into text, increasing cognitive load\added{~\cite{brown2021neural,zhaoMakingWriteConnections2025,yanXCreationGraphbasedCrossmodal2023,10.1145/3511599}}. \added{Such translation overhead not only hinders cross-modal collaboration but also restricts the creative synthesis of semantically distant concepts~\cite{10.1145/3511599} that would emerge more naturally in a truly integrated multimodal system.}

To examine this gap, we conducted a formative Wizard-of-Oz (WoZ) co-design study with 10 experienced writers to inform the design of a unified image-text multimodal system that supports the planning and translation phases of fictional story writing. We found that text and images play distinct yet complementary roles in the writing process. Writers expressed a need to directly manipulate text and images for fine-grained editing and alignment, valuing the ability to move fluidly between them.

Based on these formative results, we developed \system{}, a system for \added{experienced, LLM-literate writers} that reimagines fictional storytelling as a multimodal co-editing experience, integrating text, images, and sketches beyond traditional text-centric workflows. The design of \system{} draws on the principle of Instrumental Interaction, designing a set of instrumental operations ({lasso, collage, perspective shift, and filter}). These functions can be applied to either text or images, where a single action simultaneously affects both images and text, minimizing switching costs and preserving creative flow~\cite{beaudouin2000instrumental,shi2025brickify,AIInstrumentsEmbodyingPrompts}. Based on Dual Coding Theory, \system{} enables the alignment of text and visual representations, ensuring that edits in one modality are appropriately reflected in the other. 

We conducted a controlled study with 12 participants to examine how multimodal image–text co-editing supports fictional story writing, with a focus on evaluating the system's usability and understanding its creative support. Overall, the study showed that \system{} enhances expressiveness, immersion, and exploration, enabling participants to have more divergent ideas and write detailed narratives. Participants used the instrumental operations to refine ideas at multiple scales and explore alternative directions. While this workflow increased mental and physical workload, it also supported writers' senses of agency and ownership, as they maintained greater operational control over narrative development.

In summary, this work contributes: 
\begin{itemize}[noitemsep, nolistsep]
    \item A WoZ co-design study with 10 writers examined the practices and needs of using multimodal elements to externalize ideas and develop narratives in the planning and translating phase of fictional story writing;
    \item \system{}, a multimodal co-editing system \added{designed for experienced, LLM-literate writers} that unifies text and visual images through instrumental operations and a synchronized editing loop to support fictional story writing; 
    \item A controlled usability study with 12 participants demonstrates the potential of \system{}, suggesting that multimodal co-editing can enhance expressiveness, idea generation, and narrative development in fictional story writing.
\end{itemize}

%% file: sections/02-Related-Work.tex

\subsection{Using Visuals to Support the Cognitive Process of Fictional Story Writing}
Fictional story writing is distinct from argumentative or expository genres in its emphasis on imagination, world-building, and character development~\cite{doyle1998writer}. Writers must invent narrative worlds and characters while ensuring coherence, which poses unique cognitive challenges: abstract, nonverbal mental images must be transformed into structured narrative elements and then into text~\cite{bal2013does}. 

The Cognitive Process Model of Writing~\cite{geroDesignSpaceWriting2022, flower1981cognitive} frames writing as recursive processes of planning, translating, and reviewing. In fictional story writing, the phases from planning to translation are especially demanding, as writers move from imaginative constructs to linear verbal representation, imposing a high cognitive load due to the need for simultaneous translation and structural organization. However, visual representations can scaffold this process by externalizing abstract ideas. Research shows that picture prompts improve writing coherence~\cite{mukramah2023effect}, and visual images stimulate creativity in narrative writing~\cite{listyani2019use,han_when_2024}. In practice, sketches, maps, and diagrams externalize plot, setting, and character relationships that writers actively manipulate during planning and revision, while also serving as cognitive anchors during translation to maintain coherence and consistency.~\cite{zhaoMakingWriteConnections2025}. Dual Coding Theory~\cite{clark1991dual} explains these benefits: verbal and nonverbal systems function separately but also interact, creating richer memory traces when information is encoded in both modalities. In fictional writing, visual representations of narrative elements complement verbal planning, making abstract concepts more concrete and retrievable. When writers encounter difficulties in translation, visual anchors provide alternative access to imaginative content, reducing cognitive load and enabling more fluid expression~\cite{barsalou2008grounded}. These visual structures are not merely supportive; they function as alternative representational spaces in which writers perform cognitive operations that parallel textual editing and support non-linear narrative leaps~\cite{yanXCreationGraphbasedCrossmodal2023,10.1145/3511599}.

However, existing creativity-support systems largely leverage visuals in limited ways, focusing on inspiration, reference, or structural overview rather than enabling writers to directly manipulate visual narrative elements and align with text editing. Planning-focused systems such as CCI, Sketchar, and CharacterMeet assist authors in character and world development, through image-guided backgrounds or conversational refinement of characters~\cite{park2024character,10.1145/3677102,qin2024charactermeet}. Translation-focused tools like ScriptViz and Script2Screen aim to align textual composition with visual referents, either by retrieving reference visuals from movie databases~\cite{raoScriptVizVisualizationTool2024} or by synchronizing scriptwriting with audiovisual scene creation~\cite{wang2025script2screen}. Complexity management systems, for example, WhatIF~\cite{mishraWhatIFBranchedNarrative2025}, ClueCart~\cite{ClueCartSupportingGame}, and PlotMap~\cite{wangPlotMapAutomatedLayout2024}, help writers maintain structural coherence by visualizing branched narratives, organizing narrative clues hierarchically, or integrating spatial layouts with textual plot structures.

In these systems, writers may look at images to spark ideas, but operations such as cutting, re-ordering, or reframing narratives still have to be performed only in the verbal channel. Because the underlying story state is effectively defined only through text, visuals cannot serve as core writing operations such as restructuring events, adjusting focalization, or reorganizing character relationships~\cite{flower1981cognitive,zhaoMakingWriteConnections2025}. This separation requires writers to repeatedly convert visually grounded ideas back into verbal form for any narrative change to take effect, thereby increasing cognitive load and undermining many of the well-established benefits of external representations such as diagrams, sketches, and other forms of external cognition~\cite{kirsh2010thinking,10.1145/3511599}. As a result, images remain outside the recursive planning–translating loop. To address this gap, our work treats visual and text representations as synchronized and co-editing materials, allowing writers to manipulate narrative elements across image and text through the same set of operations and thereby more tightly aligning verbal and nonverbal with the cognitive processes of fictional story writing.

\subsection{LLM-powered Multimodality Tools for Creativity in Content Creation}
Recent multimodal creativity tools move beyond linear prompting by enabling direct manipulation of creative elements, helping creators express intentions that language alone cannot capture~\cite{AIInstrumentsEmbodyingPrompts, 10.1145/3706598.3713862}. Powered by \added{LLMs}, these systems address fundamental barriers through three complementary mechanisms. First, they externalize creative structures and support better intention expression. Tools like AI-Instruments~\cite{AIInstrumentsEmbodyingPrompts} and Brickify~\cite{shi2025brickify} transform abstract intentions into manipulable interface objects or reusable visual tokens, rendering otherwise ineffable ideas as visible, persistent, and operable elements.
Second, in recent multimodal systems, sketches always act as a nonverbal, spatially grounded modality that conveys structure, hierarchy, and relations far more efficiently than language. DrawTalking~\cite{rosenbergDrawTalkingBuildingInteractive2024} combines freehand sketching with spoken narration, enabling natural intention communication. Code Shaping~\cite{yenCodeShapingIterative2024} allows developers to make sketched annotations directly on the code editor to support fuzzy, incremental expression of intent.
Third, supporting iterative refinement,
Inkspire~\cite{lin2025inkspire} and AIdeation~\cite{wang2025aideation} accelerate variation and exploration, enabling rapid cycles of sketch-to-output or recombination of references. \added{Similarly, the AI Drawing Partner~\cite{davis2025aidrawingpartnercocreative} combines sketch recognition, text-to-image generation, and direct feedback mechanisms to enable rapid cycles of creative exploration while automatically quantifying the interaction dynamics between human and AI collaborators.}

Seeking tighter coupling for fictional story writing, recent systems push the integration of multimodal interaction in different ways~\cite{chungToytellerAIpoweredVisual2025a,chung2024patchview}.  
These systems are designed in response to growing evidence that text-only, model-driven workflows cause LLM-assisted stories to converge toward similar narrative structures, limiting exploration, reducing originality, and diminishing writers’ expressive control~\cite{10.1145/3613904.3642731,doshi2024generative,10.1145/3613904.3642625}.
WorldSmith supports layered edits and hierarchical compositions through sketches, making it easier to grow a world piece by piece instead of through single and monolithic prompts~\cite{dangWorldSmithIterativeExpressive2023}. XCreation~\cite{yanXCreationGraphbasedCrossmodal2023} supports cross-modal storybook creation by integrating an interpretable entity-relation graph, improving the usability of the underlying generative structures. Toyteller~\cite{chungToytellerAIpoweredVisual2025a} maps symbolic motions to character actions, letting users express rich social and emotional interactions that are often hard to write down explicitly only using text.
Visual Writing defines an approach where writers edit stories by manipulating visual representations to make the underlying narrative structure more comprehensible and easier to work with than linear text alone~\cite{10.1145/3746059.3747758}. 

This line of multimodal research demonstrates that combining language with gestures, sketches, and direct manipulation can offload cognitive work from linear prompting and give creators more expressive, situated channels for specifying and revising intent. Building on this line of work, our system introduces a canvas-based interface that integrates images, sketches, and text to support the externalization of mental imagery and the articulation of narrative intent.

\subsection{Instrumental Interaction}
\label{2.3}
Instrumental Interaction is central to understanding how users control and refine digital systems. Beaudouin-Lafon~\cite{beaudouin2000instrumental} proposed it as a shift from designing static interface elements to designing instruments that mediate between users and domain objects. A key principle is reification, which transforms abstract commands into persistent, manipulable objects~\cite{kent2000conceptual}. In computing, this elevates implicit system descriptions into explicit first-class entities. In LLM-assisted workflows, this principle appears in modular prompt blocks for structured edits~\cite{yen2023coladder} and in Textoshop’s reification of abstract image editing commands (e.g., tone adjustment, boolean operations, layers) into direct manipulation tools for text~\cite{10.1145/3706598.3713862}. A second principle is polymorphism, where the same instrument applies across contexts~\cite{masson2023transforming}. This reduces cognitive load by enabling predictable, transferable patterns, e.g., copy–paste works consistently across text, images, and files~\cite{angelini2015move}, and scrollbars operate similarly across documents, spreadsheets, and browsers~\cite{masson2023transforming}. Finally, reuse allows users to replay or adapt prior operations, from macros to redo commands~\cite{AIInstrumentsEmbodyingPrompts}. Systems like Spacetime exemplify this by objectifying space, time, and actions into persistent containers, enabling edits to be carried forward as manipulable entities~\cite{xia2018spacetime}. Together, these principles reduce cognitive burden by externalizing interaction histories, making them manipulable, transferable, and extensible.

In our system, we extend this perspective to the design of multimodal tools for fictional story writing. We reify narrative development as a set of instrumental operations spanning text and imagery. Through polymorphism, the same operation can be applied to both LLM-generated text and images. This combination of reification and polymorphism enables writers to shape multimodal outputs fluidly, aligning with both verbal and nonverbal perceptions.

%% file: sections/03-Formative-Study.tex
Previous research shows that multimodal tools enhance fictional story writing by making abstract concepts tangible, reducing cognitive load, and improving creativity and coherence~\cite{chungTaleBrushSketchingStories2022,chung2024patchview,dangWorldSmithIterativeExpressive2023,zhaoMakingWriteConnections2025}. However, current tools treat images as supplementary rather than integral to the creative process, leaving unclear how writers actually integrate multiple content types into a cohesive workflow.

To address this gap, we conducted a Wizard-of-Oz (WoZ) co-design study~\cite{10.1016/0950-7051(93)90017-N} examining how creators use multimodal content (images, text, sketches) when planning and drafting fictional stories~\cite{zhaoMakingWriteConnections2025,geroDesignSpaceWriting2022, flower1981cognitive}. Our investigation focused on three questions: (1) \textbf{Multimodal information use:} what types of multimodal content users employ and how they leverage these materials for idea generation; (2) \textbf{Iteration and integration:} how creators refine and combine multimodal artifacts in world-building and narrative development; and (3) \textbf{Organization of inspirations:} how creators organize, connect, and refine dispersed inspirations through multimodal manipulation. The WoZ setup simulated AI-assisted visual and textual support while sustaining the impression of an intelligent, interactive system. 

Our system is designed for writers with intermediate to expert writing expertise, rather than novices who are still learning basic narrative composition. This target group typically possesses established writing habits and a solid understanding of narrative structure. As contemporary writers increasingly incorporate LLMs into their creative workflows (idea generation, style adjustment, etc.)~\cite{10.1145/3635636.3656201,lee2024design}, we target writers who have hands-on experience using LLMs to assist their writing, even though they may not be experts in multimodal interaction or prompting.

\subsection{Process}

We designed a WoZ co-design study, positioning participants as active co-designers and treating text, sketches, and images as shared design materials~\cite{10.1016/0950-7051(93)90017-N,sanders2008co,vaajakallio2014design}.

\subsubsection{Participants}
For the formative study, we recruited 10 participants through student organizations by sharing our study announcement in group chats, each with at least two years of experience in creative writing. The group included three fictional story writers, three animation scriptwriters, two visual film creators, one new media creator, and one online fiction writer. Eight participants held a master’s degree or higher, and two held a bachelor’s degree. All participants had experience using LLMs to assist in their writing, AI familiarity ranging from casual use (fewer than two days per week, n=5) to daily workflow integration (five or more days per week, n=5).

\textit{Experimental Setup.}
Three days before the session, participants were instructed to prepare a brief fictional story outline consisting of several sentences that followed one of the narrative structures from The Seven Basic Plots~\cite{booker2004seven}, which served as the foundation for subsequent ideation and content development. \deleted{The 90-minute main session took place in either Figma~\cite{figma_home} or Miro board~\cite{MiroWebsite}, based on participant preference.}
Each session concluded with a 30-minute semi-structured interview probing how multimodal materials mediated co-creation, and what interaction patterns and workflows participants desired.

\subsubsection{Wizard-of-Oz System and Session Process}
\begin{figure*}
  \centering
  \includegraphics[width=\textwidth]{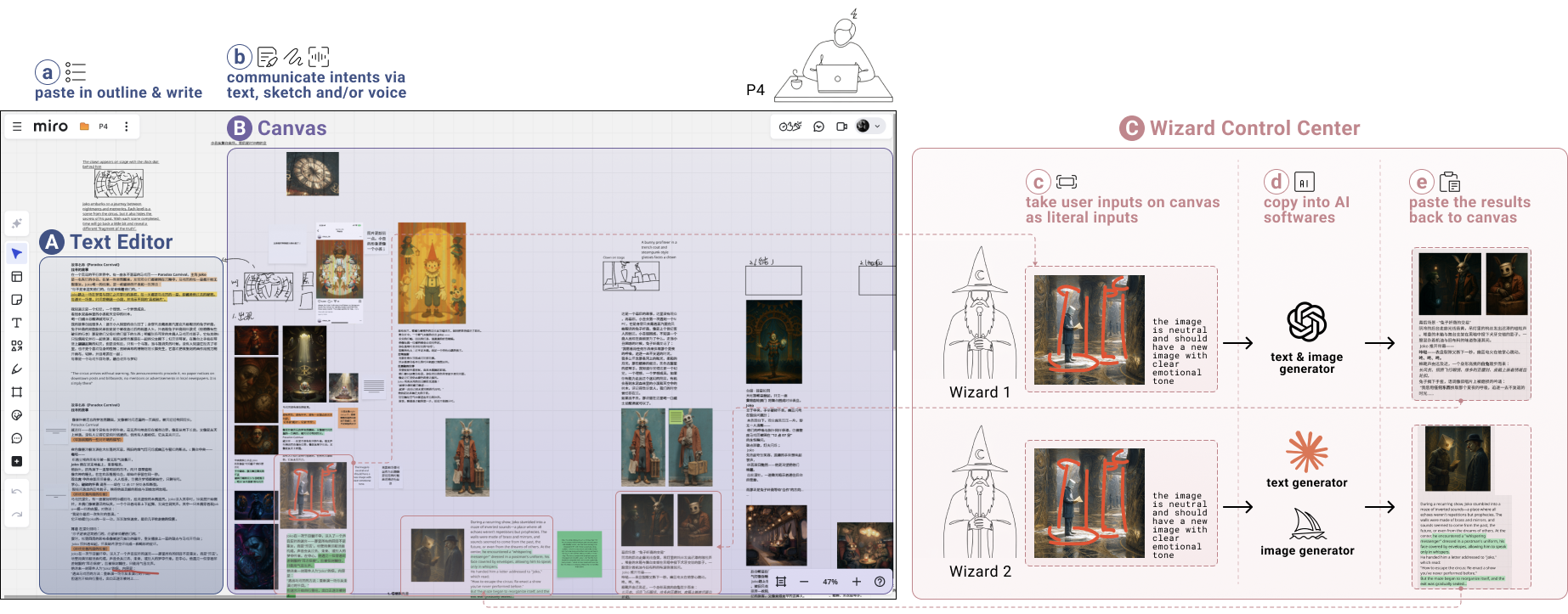}
  \caption{WoZ System and Study Process (the example of P4) (a) The user edits text in the Text editor; (b) The user writes, sketches, and speaks out their intents; (c)-(d) both wizards paste in the user's inputs into AI software windows; (e) wizards pasted the generated results back onto the canvas.}
  \label{fig:woz}
  \Description{The figure depicts the Wizard-of-Oz study setup and interaction process. It shows a text editor where the participant edits story text, a canvas where the participant writes notes, sketches, and expresses intentions, and a hidden wizard control area where researchers input user content into AI tools. The generated images and text are then returned to the canvas, illustrating the end-to-end co-design workflow.}
\end{figure*}

For the WoZ interface, we utilized the canvas in either Figma~\cite{figma_home} or Miro~\cite{MiroWebsite} as a collaborative space, which was divided into (1) a user-facing ``Text Editor'' where participants can put in the outline and edit the story, (2) the ``Canvas'' where generated images and text, and participants' notes were, and (3) a hidden ``Wizard Control Center''  (as shown in Figure~\ref{fig:woz}).
Participants communicated via voice, text (sticky notes), or hand-drawn sketches while two researchers acted as the ``Wizards (system backend)'' in real-time to generate outputs, ensuring responsiveness, copying user inputs into separate windows. Researchers ran Claude for text generation, ChatGPT (GPT-4o) and Midjourney\footnote{All models were accessed via their commercial web interfaces: \url{https://claude.ai}~\cite{claude_new}, \url{https://chatgpt.com}~\cite{openai_hello_gpt4o}, and \url{https://www.midjourney.com/}~\cite{midjourney_home} respectively, in June 2025.}, for simulating the visual engine, then pasting the results back onto the ``Canvas.'' To ensure consistency, the Wizards followed: (1) input the user's sketch/text as a literal prompt. Use the initial outline as contextual information for prompts; (2) do not offer creative suggestions unless explicitly asked; (3) to ensure diversity of output style, one researcher generated both texts and images via ChatGPT, and the other researcher generated images via Midjourney and texts via Claude. The two wizards ensured that participants were provided with results that were both timely and diverse.

\textit{Session Process.}
During the session, participants engaged in fictional story co-design through activities including generative prompts, collage, and storyboard-like arrangement while interacting with the wizards. Rather than working toward a fixed output, participants iteratively developed stories of about 300 words while envisioning how the tool itself should behave.

\subsection{Formative Study Findings}

\subsubsection{Using Multimodal Input to Reify Vague Ideas}\label{Formative_multimodal}

As shown in the Appendix~\ref{tab:formative_grouped}, participants utilized multimodal expressions, including text, sketches, and images, as co-design materials to articulate and negotiate intentions with the wizarded system. 

\textbf{Sketches} externalized vague intentions and spatial imagination. For example, P4 envisioned a scene where clown Joko appeared on stage and created a sketch with textual annotations describing the intended atmosphere, hoping AI could elaborate narrative details. Sketches were also frequently used to express ideas that participants found difficult to convey through prompts alone (P6, P7, P8), as they were seen as carrying richer layout and spatial information. \textbf{Text} functioned as the primary medium for conveying intent, allowing participants to express connections between desired content and existing stories. Participants also frequently used textual annotations to specify story parts or image types for AI-generated content and to guide the direction of content generation. They also used textual annotations to record the inspirations they received and how those ideas might be used in writing. \textbf{Images} communicated style and mood expectations. P3 altered an AI-generated picture's style by supplying a reference image, while P1 noted, \textit{``If possible, I want to use a `supporting image'---a vague reference picture---as a basis, expecting the AI to generate more detailed images derived from it.''} \deleted{Combining image outputs with textual descriptions helped participants enrich their limited knowledge. For example, P7 requested AI-generated designs of an ancient Chinese poison bottle as a narrative element, noting her vague understanding of the concept.} In addition, sketches were often paired with images or text to further articulate the intentions participants held in mind (P4, P7).
\deleted{For the content, participants most often sought LLM elaboration on characters, objects, and scenes, expressing the need for assistance with character design refinement, setting depictions, or object visualization.} 

These findings highlighted the value of systems that accept different modality input and help co-designers transform nascent ideas into concrete narrative materials.

\subsubsection{Image-Text Interplay as Complementary Design Moves}\label{Formative_Text–Image Interplay}

Participants perceived text and images as distinct yet complementary, and we observed them often switching between abstraction (text) and concreteness (images) as a recurring co-design pattern.

\textit{Text as open imagination.} Participants described text as a ``blank canvas'' for boundless imagination. P4 noted, \textit{“Text allows me to imagine many things in my mind,”} and P1 emphasized text as ``infinite imagination on a blank page.'' P8 highlighted that text helped set up the narrative structure before layering in visuals. This suggests systems should treat text as a flexible space for ideation and intent communication, where ambiguity can be preserved rather than early resolved. 

\textit{Images as concreteness, inspiration, and feedback.} Images grounded abstract ideas, while their randomness often sparked unexpected inspiration. P1 explained: \textit{``The randomness in AI-generated images goes beyond what I want or can express; it helps me imagine the next step of the story.''} Similarly, \deleted{P4 refined Lily’s behavior based on an unexpected visual detail, and P3 used images as feedback for progressive refinement. }P10 regarded referring to images as a ``look-and-write exercise'' that scaffolds scene construction. These accounts highlight the value of image outputs not just as illustrations but as provocations. \deleted{Participants leveraged it through image-based iteration, selection, and reinterpretation.} 

\textit{Complementary interplay.} Participants emphasized that neither modality sufficed alone: images ``set the vibe,'' while text reframed meaning. P8 noted, \textit{``Images can serve as references for appearance when I don't have many ideas, while text quickly triggers associations.''} We observed participants iteratively moving between text for open-ended imagination and images for concrete grounding, forming a cycle of divergence and convergence (P4, P5, P8, P9, P10). Four participants (P1, P2, P5, P7) also expressed a desire for text and image changes to be synchronized, so they would not need to constantly cross-check and compare updates across modalities. In addition, the concurrent presentation of text and images further facilitates narrative expression. As P7 explained, \textit{`` \added{When text and images appear together, I can describe the scene while referencing the image and directly adopt AI-generated text when it proves useful. This also allows inspiration to emerge.}''}  \deleted{They noted that when an image is edited, the corresponding text should update accordingly.}

This interplay suggested design opportunities for systems that incorporated smooth transitions between text and images while aligning these two modalities, enabling participants to fluidly move between abstract exploration and concrete elaboration

\subsubsection{Direct Manipulation of Multimodal Artifacts}\label{Formative_Direct_Manipulation}

Participants expressed a strong interest in treating text and images as manipulable, recombinable design materials. Two recurring practices pointed to design needs for more direct multimodal manipulation.

\textit{Collaging and Recombination.} Participants frequently merged elements across outputs to spark new ideas. P9 envisioned combining ``the house from the first AI-generated images with the street from the second picture'' to construct scenes, while P7 highlighted that ``randomly combining characters and scenes'' could inspire unexpected connections when accompanied by textual descriptions. \added{She also expressed that visualizing hard-to-imagine scenes makes it easier to write richer narrative descriptions.}
 Such practices illustrated the potential of collage and recombination as creative strategies.

\textit{Granular Editing and Annotation.} Beyond recombination, participants desired fine-grained control over outputs. They use sticky notes to capture details for iteration and to serve as prompts for later development. Participants also wanted more localized operations, such as regenerating specific regions (P1, P3), extracting and reusing circled image elements (P5), or annotating character personas for refinement (P2). \deleted{They left narrative prompts for later translation, e.g., P2’s note \textit{``Ending could be related to why this postman job even exists.''}} These behaviors emphasized editing and annotation as both vehicles for iteration and a bridge to subsequent writing. 

Together, these findings suggested systems should enable flexible recombination, localized editing, and traceable annotations to help creators iteratively refine narrative materials.

\subsubsection{From Fragmented Inspirations to Coherent Storylines}\label{Formative_Fragmented}

While participants often highlighted text or circled inspiring image details, organizing these dispersed fragments into coherent narratives was a persistent challenge in the planning phase. As P2 noted, \textit{``Everything quickly became too messy on the canvas,''} and P4 likened fragments on the canvas to ``many cards that required connections,'' where narrative coherence depends on linking passages, characters, and settings from scattered parts. Furthermore, P6 wished for mind map–like tools to scaffold this process. Participants also requested clustering notes, surfacing latent relations (e.g., by character/object/setting), and consolidating materials into reusable ``setting cards'' to ensure cross-chapter consistency and avoid logic conflicts (P4, P7). 

These challenges pointed to opportunities for systems that transform fragmented inspirations into structured storylines by supporting clustering, relation mapping, and the creation of reusable narrative units that preserve coherence across iterations.

\subsection{Design Goals} 
Drawing on insights from our WoZ co-design study, prior work on multimodal LLM tools, and theories of Structural Mapping and Instrumental Interaction (Section~\ref{2.3}), we identify four design goals for a multimodal content creation interface that supports the planning and translating phase in fictional story writing~\cite{geroDesignSpaceWriting2022, flower1981cognitive, zhaoMakingWriteConnections2025}.

\begin{figure*}[htbp]
  \centering
  \includegraphics[width=\textwidth,keepaspectratio]{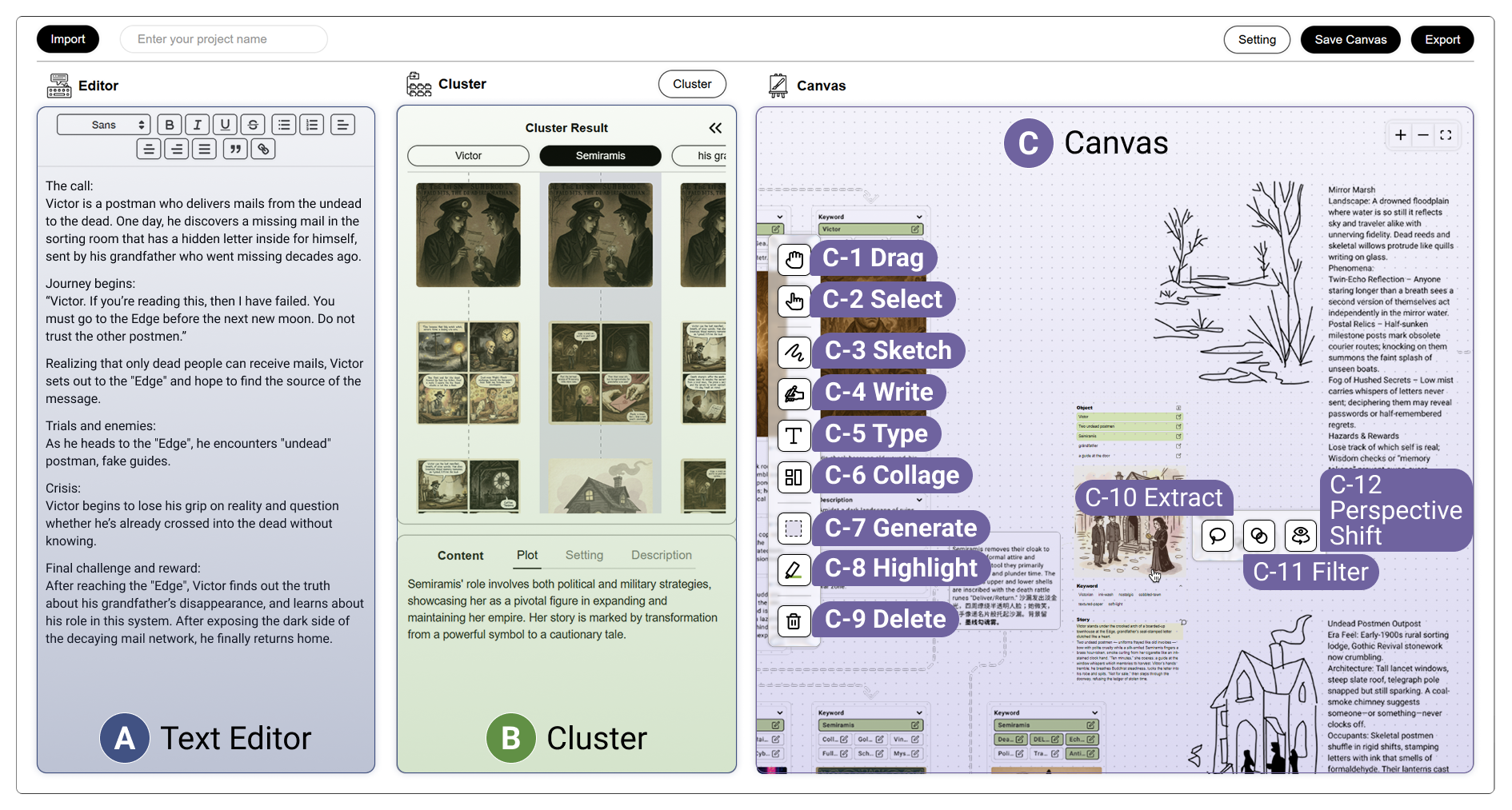}
  \caption{Vistoria’s interface: a left Text Editor, a central Cluster panel (can be collapsed when not used), and a right free-form Canvas.}
  \Description{The figure presents Vistoria’s interface layout. It consists of a left-side Text Editor displaying the current story draft, a central Cluster panel that aggregates highlighted narrative elements, and a right-side free-form Canvas for sketching, placing images, writing text, and performing image–text co-editing operations.}
  \label{fig: global layout}
\end{figure*}

\begin{itemize}
    \item \textbf{DG1: Supporting the Expression of Ideas through Multimodality.}
    Grounded in the findings in Section~\ref{Formative_multimodal}, our system should provide multimodal mechanisms combining sketches, text, and images to capture early intention, imprecise expressions, and help transform them into concrete narrative materials for further iterative editing or re-organizing.

    \item \textbf{DG2: Aligning Text and Images for Iterative Creative Exploration.}  
    Informed by findings in Section~\ref{Formative_Text–Image Interplay}, our system should enable fluid cross-modal iteration: textual edits can be re-visualized, and image refinements can inform text descriptions.
    Grounded in Dual Coding Theory, text and image updates should also be synchronized to more effectively align verbal and nonverbal perception.
    
    \item \textbf{DG3: Enabling Polymorphic Cross-Modal Manipulation.}  
    Informed by findings in Section~\ref{Formative_Direct_Manipulation}, our system should support direct manipulation interactions for both text and images. Guided by Instrumental Interaction’s principle of polymorphism, we should design the same instrument for cross-modal editing to reduce switching costs and enable writers to manipulate textual and visual fragments while maintaining narrative coherence.

    \item \textbf{DG4: Organizing and Reusing Fragments into Coherent Narratives.}  
   Informed by the findings in Section~\ref{Formative_Fragmented}, our system should support clustering and organizing fragments and fleeting ideations during the exploratory phase, surface latent connections, and consolidate dispersed inspirations into coherent, evolving narrative structures that support translation into final writing. 
\end{itemize}

%% file: sections/04-system.tex
In this section, we present the key features of \system{}.  As shown in Figure~\ref{fig: global layout}, the interface comprises three primary components: a left Text Editor, a central collapsible Cluster panel, and a right Canvas interface. The Text Editor displays the current story draft, serving as contextual information for content generation. The right Canvas supports freeform sketching, text input, and image-text generation and editing tools. The central Cluster panel aggregates highlights and annotations from Canvas,  displaying related plots, settings, and descriptions of each highlighted element for easy reference and overview.

\begin{figure}[htbp]
  \centering
  \includegraphics[width=\columnwidth,keepaspectratio]{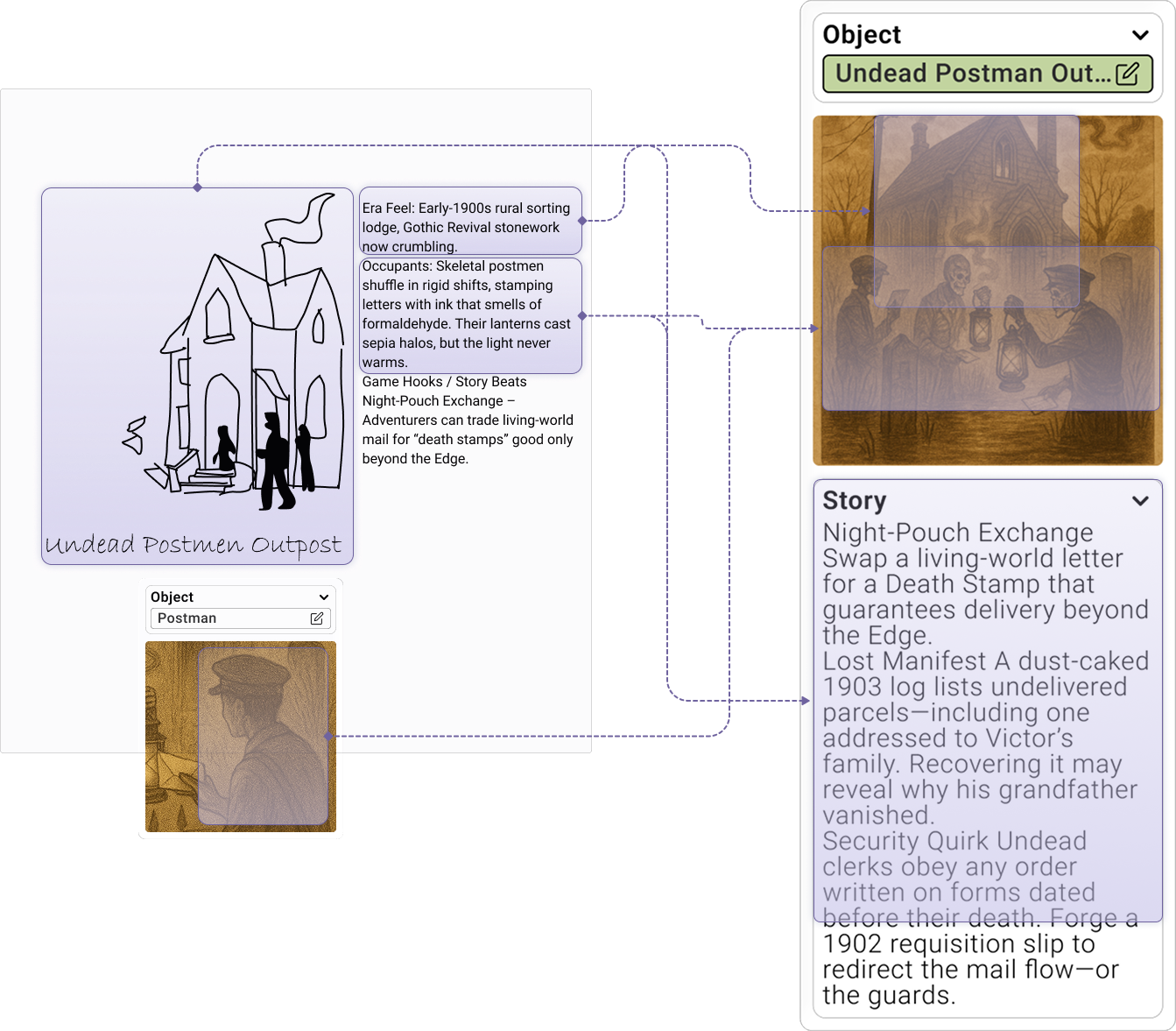}
  \caption{\system{} analyzes selected text, sketches, images, and broader story context to generate image–text cards that externalize narrative development directions.}
  \label{fig: Feature - Generation}
  \Description{The figure shows how Vistoria generates image–text cards from multimodal input. Selected regions containing text, sketches, and images on the Canvas are analyzed together with the broader story context. The system then produces image–text cards that externalize potential narrative development directions.}
\end{figure}

\begin{figure*}[t]
  \centering
  \includegraphics[width=\textwidth,keepaspectratio]{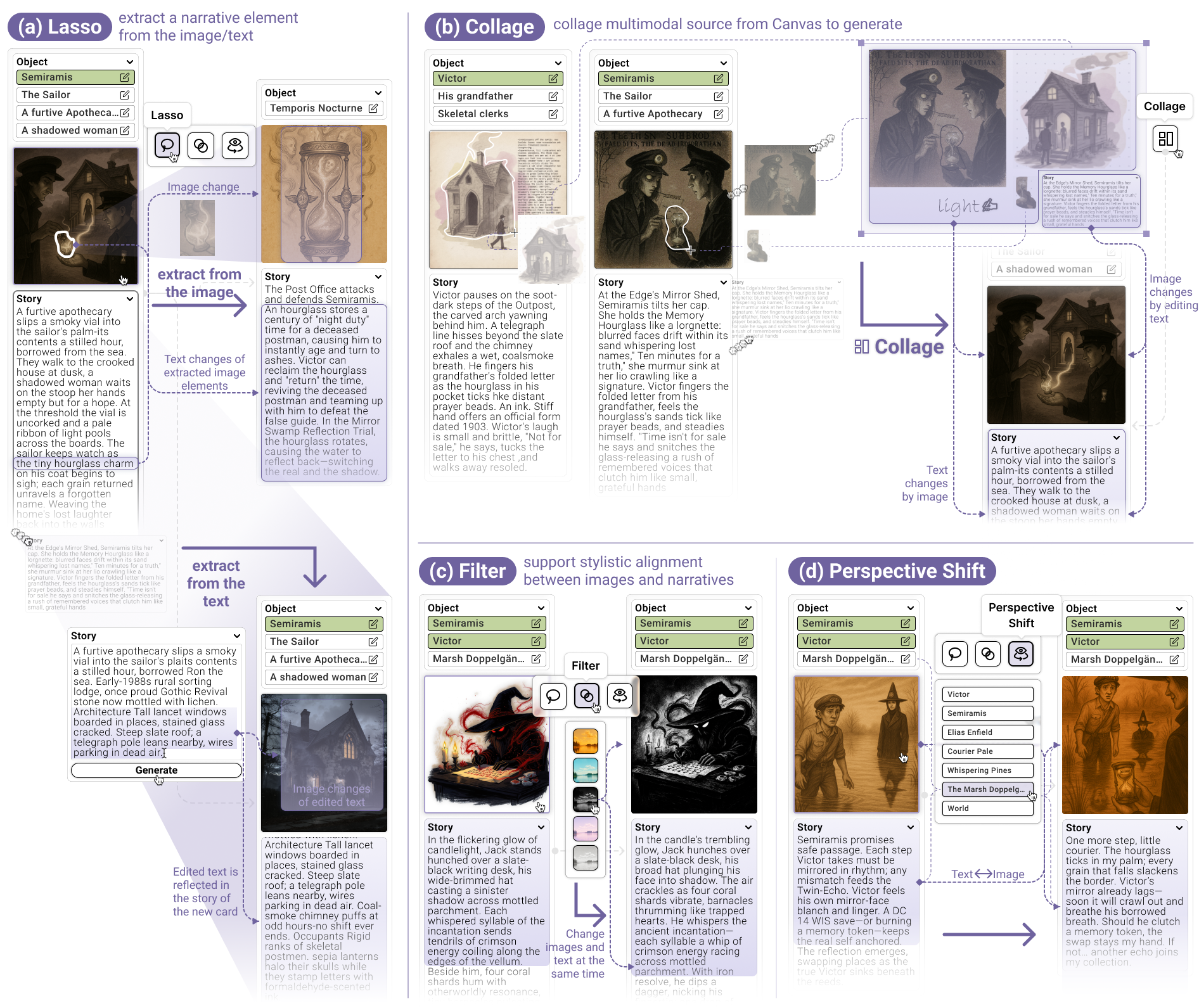}
  \caption{A set of instrumental operations for image-text co-editing to enhance planning and translating of fictional story writing: (a) Lasso selects regions for coupled image–text edits. (b) Collage enables writers to extract elements and compose across cards to discover new narrative directions. (c) Perspective Shift changes an image’s viewpoint and automatically regenerates the story’s point of view (first/third/second person). (d) Filters align visual style and textual tone (e.g., melancholic/dreamy) by jointly altering image effects and rewriting prose.}
  \label{fig: Feature - Image Editing}
  \Description{The figure illustrates four instrumental operations for synchronized image–text co-editing. The Lasso operation selects localized regions for focused edits, the Collage operation combines extracted elements across cards, the Filter operation aligns visual style with textual tone, and the Perspective-Shift operation changes visual viewpoint and narrative voice together. Each operation affects images and text simultaneously.}
  
\end{figure*}

\subsection{Key Features}



\subsubsection{Reifying Intention through Multimodal Generation}  

The system enables writers to externalize early, vague ideas using multimodal inputs (DG1). To support this, \system{} transforms multimodal input (sketches, text, and images) into cards that pair a representative image with its corresponding narrative segment. Writers select a region of interest, capturing all multimodal elements within that area. \system{} then infers from these multimodal elements and broader story context to generate image–text cards that reflect the writer's intention for narrative direction (Figure~\ref{fig: Feature - Generation}). These cards serve a dual purpose: they transform ephemeral multimodal input into persistent, regenerable artifacts on the Canvas, while also functioning as semantically aligned units where image and text convey the same underlying meaning.

\system{} further balances precision and exploration by offering two complementary generation modes. In \emph{Exact Craft} mode, single cards closely adhere to the author’s expressed intention to concretize specific ideas. In \emph{Creative Spark} mode, three cards are generated to represent diverse options based on the writer's intention. The system deliberately introduces variation around characters, settings, or objects, providing alternative prompts that can inspire new directions.

\begin{figure*}[b]
  \centering
  \includegraphics[width=\textwidth,keepaspectratio]{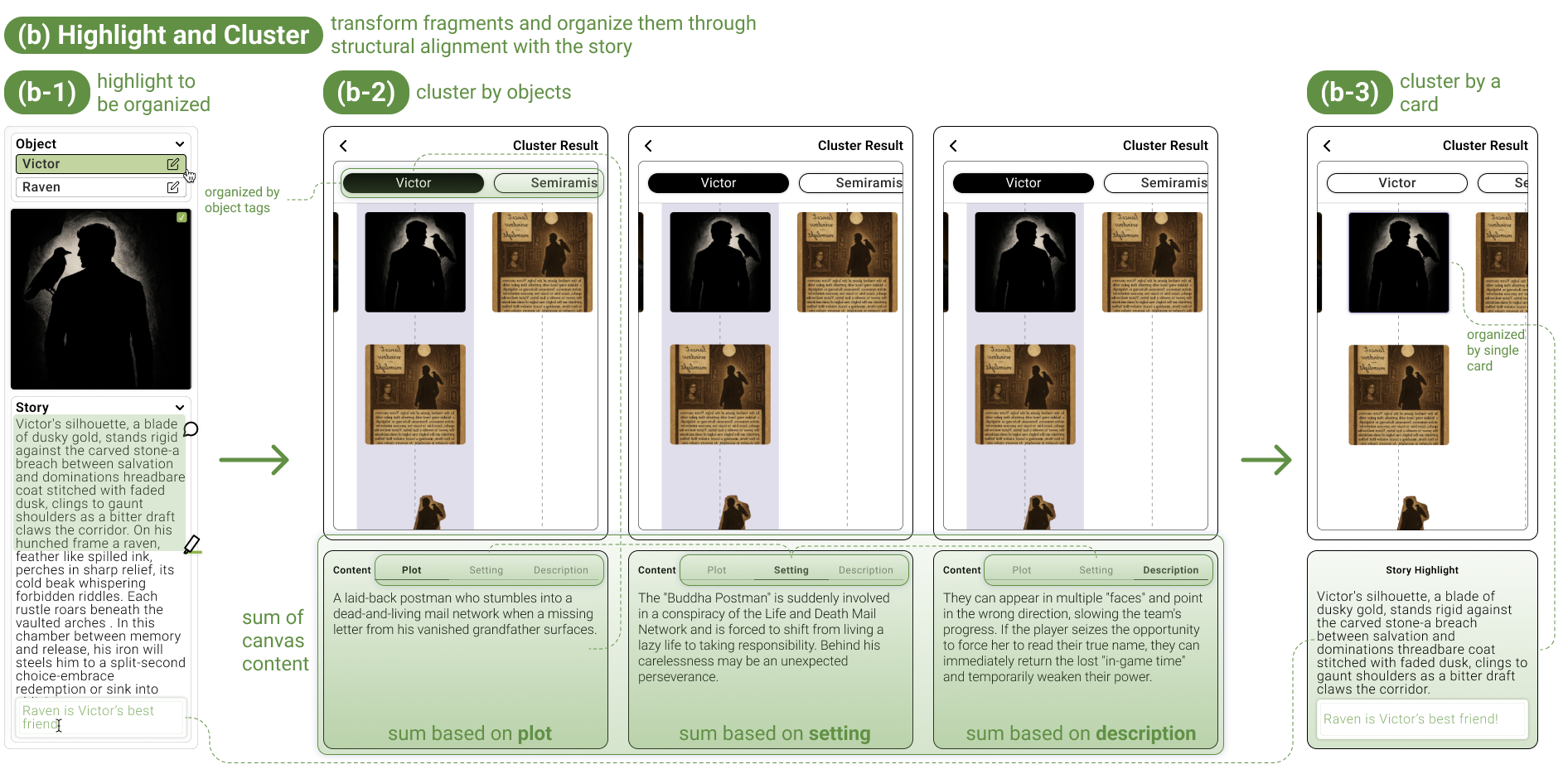}
  \caption{Writers highlight objects and text segments on cards (b-1); the Cluster panel aggregates these by character/object/scene and can auto-summarize settings/plot/description about a certain object to guide final writing (b-2); Clicking on a specific image reveals the corresponding highlights and comments from earlier phases left on Canvas (b-3).}
  \label{fig:Cluster}
  \Description{The figure shows how highlighted text segments and objects on cards are aggregated in the Cluster panel. Narrative elements are grouped by character, object, or scene, and the system generates structured summaries of plots, settings, or descriptions. Selecting an item reveals its associated images, text highlights, and comments to support coherent story construction.}
\end{figure*}

\subsubsection{Synchronized Image–Text Co-Editing through Instrumental Operations}  

Fictional story writing benefits from fluid movement between abstract textual reasoning and concrete visual imagination. The system should tightly align text and images so that edits in one modality fluidly inform the other (DG2, DG3). To address this, we introduce a set of \emph{Instrumental Operations} (Figure~\ref{fig: Feature - Image Editing}) designed around three principles: (1) \added{}\emph{Reification (instrumental interaction)}, which draws on familiar image-editing operations to make abstract image–text co-editing actions more concrete and manipulable; (2) \emph{Polymorphism (instrumental interaction)}, designing a set of instrumental operations (Lasso, Collage, Filter, Perspective Shift) which ensure the same operations apply uniformly across text and images to lower switching cost; 
and (3)
\emph{Dual Coding Theory}, which indicates that verbal and nonverbal changes should be aligned to maintain coherence across cognitive channels.

\paragraph{\textbf{\textit{Lasso.}}}  
The \textit{Lasso} instrument exemplifies reification by turning the abstract action of ``focusing on part of a story'' into a manipulable unit: selecting a region in \textit{either} an image or a fragment of text triggers the generation of a new card focusing on the selected part with enriched narrative and visual details. 
Through \textit{polymorphism}, the same selection logic applies across modalities---whether circling a visual detail or isolating a text segment---providing a consistent interaction pattern. The text within the selected area in the original content will be emphasized to form a new card. The extracted portion of text is used to regenerate the corresponding image. The lassoed image region and the expanded story fragment correspond to one another, aligning visual and textual perspectives within the same narrative unit (Figure~\ref{fig: Feature - Image Editing} (a)).

\textbf{\textit{Collage.}}  
The \textit{Collage} instrument reifies the abstract act of ``recombining inspirations'' into a tangible manipulation: fragments of images, sketches, or text can be directly composed within a collage frame to form a new card. The same cut–paste–combine logic applies uniformly across modalities---an image region, a text excerpt, or a sketch element can all be treated as compositional materials for intention-based generation.  The system interprets the spatial arrangement of these multimodal pieces as narrative intent, generating a card where textual descriptions and visual depictions are aligned. For instance, merging two character fragments not only produces a combined image but also generates a new story segment situating them together, ensuring that narrative and imagery evolve in sync (Figure~\ref{fig: Feature - Image Editing} (b)).

\textbf{\textit{Filter.}}
Stylistic coherence is critical in fictional story writing, as consistent affective and aesthetic cues sustain narrative transportation~\cite{green2000role}, activate readers’ interpretive schemas~\cite{arbib1992schema}, and enhance the emotional resonance of literariness~\cite{miall1999literariness}. In \system{}, the \textit{Filter} instrument reifies this abstraction into a concrete tool: applying a ``melancholic'' or ``dreamy'' filter adjusts the visual style and rewrites the accompanying prose to match the emotional tone (Appendix Table~\ref{tab:filters}). Through \textit{polymorphism}, the same filter operation works seamlessly across modalities, leveraging the correspondence between visual style in images and emotional tone in text to simultaneously act on both. By making intangible stylistic intentions manipulable and synchronized, filters expand expressive possibilities while maintaining narrative immersion (Figure~\ref{fig: Feature - Image Editing} (c)).

\textbf{\textit{Perspective Shift.}}
Fictional story writing often utilizes perspective shift, and narratology highlights that changes in voice and focalization fundamentally reshape how events and characters are perceived~\cite{genette1980narrative}. Cognitive poetics further shows that such shifts alter readers' empathy and immersion. First-person narrations foster intimacy, while third-person perspectives enable broader structural awareness~\cite{keen2007empathy}. The \textit{Perspective-Shift} instrument reifies this narratological concept into an actionable operation: changing the visual viewpoint of a scene automatically regenerates the story fragment from a first-, third-, or second-person perspective. Through \emph{polymorphism}, this instrument applies consistently across modalities, altering either an image or its accompanying text triggers a corresponding adjustment in the other. The shift carries the same meaning across text and image: a new camera angle in the image corresponds to a new narrative voice in the text, allowing writers to explore empathy, distance, and awareness in a synchronized manner (Figure~\ref{fig: Feature - Image Editing} (d)).


\subsubsection{Highlight Elements and Cluster.}
Writers often struggle to integrate scattered highlights and annotations on the Canvas into coherent storylines, leaving ideas fragmented across cards (DG4). \system{} addresses this by transforming the dispersed fragments into reusable narrative building blocks, aligning them structurally across characters, objects, and scenes. This process is centered in the Cluster panel (Figure~\ref{fig:Cluster}), which turns fragmented inputs into organized knowledge assets.

On the Canvas, writers can highlight textual segments, edit stories directly, and add inline comments noting potential uses in later drafting. Story objects, such as characters, settings, or scenes, are represented as editable keywords that can be highlighted by themselves. The system automatically links each highlighted object to its associated text, consolidating references across multiple cards.

The Cluster panel then aggregates all highlighted objects into an organized overview of evolving narrative elements. This eliminates the need for manual scanning of scattered cards and provides writers with a dynamically updated, object-centered workspace. Selecting an object reveals its complete set of associated materials, including linked images, highlighted text segments, and comments, which creates a multimodal, context-rich reference for downstream writing. Beyond simple aggregation, the panel supports higher-level knowledge construction through its summary feature. When this feature is invoked, the system generates structured summaries of settings, descriptions, and plot elements derived from highlights and comments. These summaries distill fragmented annotations into narrative building blocks~\cite{burroway2022writing}, enabling writers to iteratively scaffold coherent storylines from previously disjointed ideas.

\subsection{Implementation}

We adopted a decoupled front-end/back-end architecture. The React\footnote{\url{https://react.dev/}} front-end enables efficient rendering for complex interactive interfaces, while the Flask\footnote{\url{https://flask.palletsprojects.com/}} back-end flexibly handles model calls. Axios manages asynchronous communication between layers. 

\subsubsection{Front-end.}





The front-end consists of three main modules: the Canvas, the Cluster, and the Text Editor. Zustand\footnote{\url{https://github.com/pmndrs/zustand}} centrally manages the global state (including canvas nodes, cluster selections, and text content) to ensure consistency across all modules. To protect privacy during user studies, per-session data is stored in sessionStorage and automatically clears when the tab closes, while users can manually export Canvas nodes and text content via the top toolbar.

\textit{Canvas Module.} The right-side Canvas module consists of four distinct layers: 
(1). Node Interaction Layer (Bottom): This layer uses React-Flow\footnote{\url{https://reactflow.dev/}} to maintain a dynamic node-edge graph. Node types include card, collage, text, sketch, handwriting, and image. All nodes share basic properties, such as ID, type, and coordinates, for edge linking, but the internal data structures of those node types vary for rendering. For example, the ``card'' node includes features such as image modification tools, image lasso selection, image highlighting, object manipulation, and basic information display. 
(2). Pen-based Input Capture Layer: This layer supports natural interactions like freehand sketching, writing, and lasso selection. It uses the perfect-freehand library\footnote{\url{https://github.com/steveruizok/perfect-freehand}} to smooth captured points and convert them into Scalable Vector Graphics (SVGs). Each captured SVG stroke is added to the graph as a new node. 
(3). Generation Selection Layer: A Document Object Model (DOM) based screenshot function ensures visual and positional consistency. After clicking generate, the front-end sends both the screenshot and structured data of the nodes inside the selected area to the backend, adds a new node to the graph, and awaits the returned data. (4). Tool Layer (Top): This is the most visible layer. It contains operation tools and Canvas control tools.

\textit{Cluster Module.} The middle module displays selected information, including objects, images, text, and annotations. Users filter this content from the node-edge graph by selecting specific objects. A button allows users to expand or collapse this entire area.

\textit{Text Editor Module.} The left-side Text Editor uses the React-Quill\footnote{\url{https://github.com/zenoamaro/react-quill}} component. It provides lightweight rich-text editing capabilities designed to align with the narrative structure.

\subsubsection{Back-End Multi-Agent Flow.}
The backend is organized as a multi-agent pipeline, where each prompt-specialized LLM agent is organized sequentially. Rather than relying on a single model, the system decomposes the workflow into three cooperating functional agents:

\textit{Narrative Construction Agent.}
This agent takes multimodal context---such as Canvas screenshots which incorporate all information (sketches, text inputs, and images) contained within the user-selected region to produce structured textual outputs---including user intentions and story segments. Its prompts enforce consistency with contextual information such as the existing story content in the Text Editor and the global stylistic constraints. In essence, the agent transforms the user’s multimodal inputs into coherent narrative storylines. During implementation, the GPT-o4-mini\footnote{%
We used OpenAI's o4-mini model via the OpenAI API (model ID \texttt{o4-mini}) in August~2025~\cite{openai_o4_mini_api}.} is used to process Canvas screenshots and contextual information to infer user intentions because of its rapid inference speed and strong reasoning ability. GPT-4o\footnote{%
We used OpenAI's GPT-4o model via the OpenAI API (model ID \texttt{gpt-4o}) in August~2025~\cite{openai_gpt4o_api}.} is used to refine these inferred intentions from o4-mini and contextual information into polished, coherent story segments. Prompts for precise description generation of o4-mini are shown in Appendix~\ref{Prompts}). When applying instrumental operations for editing, the story segment from the previous card is also read in prompts and modified accordingly.

\textit{Visual Synthesis Agent.}
Using the narrative produced by the Narrative Construction Agent as prompts, this agent supports image generation using either the GPT-4o API or the FLUX diffusion model\footnote{%
We used the FLUX.1 diffusion model via the Black Forest Labs API in August~2025~\cite{bfl_flux1_api}.}. When reference images or screenshots are available—such as during instrumental-operation edits that modify images or when multimodal inputs include a base image or sketches—GPT-4o is used for image generation, leveraging its strong capabilities in image understanding and re-generating based on base images. In all other cases, the FLUX diffusion model is used due to its faster generation speed. 

\textit{Memory Agent.}
The backend maintains a persistent, globally accessible memory of all previously generated image-text pairs. This agent coordinates read/write operations to this store, enabling multi-turn reuse, cross-scene integration, and contextual continuity. When the user later requests to merge scenes, change details, or perform local edits, this agent retrieves relevant image-text pairs and passes them back to the earlier agents to re-initiate the pipeline.

%% file: sections/05-User-Study.tex
To evaluate the usability of \system{} and to understand how these multimodal interactions design support creativity, we conducted an exploratory lab user study with 12 participants. 
We structure our user study around two complementary components:

(1) A usability evaluation of \system{} aiming to answer the following research questions: 

\textit{\textbf{RQ1:} How useful are the multimodal co-editing functions, and in what ways do they influence participants’ workflows?}

\textit{\textbf{RQ2:} How does multimodal image–text co-editing affect participants’ workload compared to a text-only baseline?}

(2) A pilot study examining the creativity support provided by \system{}, focusing on:

\textit{\textbf{RQ3:} How does multimodal image–text co-editing influence participants’ ideation and the development of fictional stories?}

\textit{\textbf{RQ4:}  How does multimodal image–text co-editing influence participants' sense of agency and ownership?}

RQ1 evaluates the usefulness of these multimodality functions and the strategies participants adopted in relation to DG3 (polymorphic instrumental operations).
RQ3 explores how DG1 (reifying intentions through multimodal input) and DG2 (image-text alignment) support creative ideation and narrative development.
The design of the cluster panel corresponding to DG4 functions mainly as auxiliary support and is not central to our user evaluation of multimodal interaction.

Note that we used the sense of ownership to refer to the writer’s ``sense of possession'' over the resulting narratives within the system, even the AI-generated artifacts\added{~\cite{10.1145/3706598.3713522, lee2024design, draxler_ai_2024, 10.1145/3639701.3656325}}. The declaration of the sense of agency, on the other hand, refers to the writer’s awareness of ``initiating, executing, and controlling'' key actions in the writing and artifact editing process~\cite{moruzzi_creative_2022, legaspi_synthetic_2019, lee2024design}.

\subsection{Participants}
As the system targets intermediate to expert participants who understand narrative structure and already use LLMs in their creative workflows, we recruited 12 participants (6 males, 6 females, aged 21–32, M=25.5), all with prior creative writing experience. Participants were also recruited through student organizations by sharing our study announcement in their group chats. All held Bachelor's degrees, with backgrounds spanning science, arts, design, or communication. Their creative practices included fiction writing, screenwriting, songwriting, advertising, and media production. Participants' creative writing experience ranged from under one year (n=5) to over seven years (n=1), with others reporting 1–3 years (n=3) or 4–7 years (n=3). All participants were familiar with LLMs (e.g., ChatGPT, Gemini, Claude) and had used them for idea generation, editing, descriptive support, content expansion, world-building, and style imitation. Among the 12 participants, 6 use LLMs daily, while the remaining 6 are evenly split across several times a week, occasionally, and rarely (2 in each).
Each participant received a \$40 USD compensation after finishing the experiment.

\subsection{Procedure}
\subsubsection{Apparatus}
Sessions were conducted on a laptop computer with keyboard and mouse for typing, dragging, and selecting. To support sketch input, we provided an external tablet (iPad) for freehand drawing on the canvas. 

The baseline condition presented a side-by-side interface with a text editor and GPT-4o~\cite{openai_hello_gpt4o} conversational panel, enabling both manual editing and LLM-assisted text/image generation. Participants completed two story-writing tasks (Appendix~\ref{Topic}), each extending a given story beginning into a 300–500 word draft. Tasks were counterbalanced across conditions (Baseline vs. \system{}).

\subsubsection{Study Procedure}
The study followed a within-subjects design with counterbalanced condition order. After informed consent and a demographic survey, participants were introduced to \system{} through a written guide and tutorial video, followed by a short hands-on exploration (15 minutes).

In each condition, participants first focused on world-building and idea exploration (20 min) and then on refining and improving the story (20 min). We divided the writing task into two phases (exploration and refinement) to prevent participants from prematurely committing to a single storyline and to reduce fixation, thereby encouraging broader ideation before focused improvement~\cite{suh2024luminate}. 

After each condition, participants completed surveys including NASA Task Load Index (TLX)~\cite{hart1988development} and Creativity Support Index (CSI)~\cite{cherry2014quantifying}. These surveys were chosen to be consistent with the standard measures employed in previous HCI system work on multimodal LLM-assisted ideation and storytelling~\cite{choi2024creativeconnect, 10.1145/3746059.3747772, darejeh2024criticalanalysiscognitiveload}.
All surveys used 7-point Likert scales. After both conditions finished, participants also completed a 15–20 minute semi-structured interview. All sessions were video-recorded via Zoom. We collected system logs, final story drafts, canvas artifacts, image–text pairs, and interview transcripts for analysis.

\subsubsection{Data analysis}

We employed a mixed-methods approach to systematically analyze three types of data.

For the qualitative interview data, we conducted an inductive, grounded theory-informed analysis~\cite{thornberg2012informed}. First, two authors independently performed open-coding on 33\% of the data, generating an initial set of 30 distinct codes. The coders then met to compare code applications, resolve discrepancies through negotiated agreement, and refine the wording and boundaries of each code. Through several rounds of discussion, they reached full consensus on all coded segments and consolidated the initial codes into a shared codebook. Using the refined codebook, the two authors independently coded half of the remaining transcripts, meeting regularly to prevent coding drift and to determine whether newly emerging codes should be incorporated. Ongoing constant comparison within and across interviews was used to further refine relationships between codes. Finally, we clustered the codes into four higher-level themes that map onto our design goals. The final codebook is shown in Appendix~\ref{codebook}.

 Second, interaction data were analyzed through structured video coding by two authors to quantify tool usage frequency and modality switching events, and researchers aligned them with system logs on an event-by-event basis. Finally, for survey measures, we conducted paired-sample t-tests under the assumptions of normality and homogeneity of variance. When assumptions were violated, we used the Wilcoxon signed-rank test. Given the sample size limitations, we treat these quantitative results as descriptive signals intended to triangulate with and support the qualitative themes.

\section{Study Results}\label{study_results}

\begin{table*}[t]
\centering
\small
\caption{Comparison of survey results: \system{} vs. Baseline. Sig.: * $p<.05$; ** $p<.01$}
\label{table:Statistics}
\setlength{\tabcolsep}{6pt}
\renewcommand{\arraystretch}{1.15}
\begin{tabular}{llcccccc}
\toprule
& & \multicolumn{2}{c}{\system{}} & \multicolumn{2}{c}{Baseline} & \multicolumn{2}{c}{Statistics} \\
\cmidrule(lr){3-4}\cmidrule(lr){5-6}\cmidrule(lr){7-8}
& & mean & std & mean & std & p & Sig. \\
\midrule

\multirow{6}{*}{NASA-TLX}
& Mental          & 5.16 & 1.528 & 3.167 & 1.337 & 0.0000   & ** \\
& Physical        & 4.667 & 1.723 & 2.083 & 0.669 & 0.0002   & ** \\
& Temporal        & 2.917 & 1.443 & 2.750 & 1.215 & 0.7723   & -- \\
& Effort          & 4.083 & 1.443 & 3.500 & 1.834 & 0.3388   & -- \\
& Performance     & 5.250 & 1.712 & 5.083 & 1.564 & 0.7986   & -- \\
& Frustration     & 2.750 & 1.183 & 1.750 & 0.622 & 0.0204   & * \\
\midrule
\multirow{6}{*}{\parbox[t]{3.1cm}{Creativity Support\\ Index}}
& Exploration           & 4.917 & 1.240 & 4.750 & 1.485 & 0.7126 & -- \\
& Expressiveness        & 6.083 & 0.996 & 4.333 & 1.775 & 0.0232 & *  \\
& Immersion             & 4.917 & 1.505 & 2.750 & 1.545 & 0.0006 & ** \\
& Enjoyment             & 5.333 & 1.435 & 4.917 & 1.379 & 0.1753 & -- \\
& Results Worth Effort  & 5.250 & 1.357 & 5.583 & 0.793 & 0.5166 & -- \\
& Collaboration         & 5.500 & 0.674 & 4.583 & 1.505 & 0.0418 & *  \\
\bottomrule
\end{tabular}
\end{table*}



\subsection{The Usability of \system{}}

To address RQ1 and RQ2, we analyzed how participants engaged with the designed instrumental operations (\textit{Lasso, Collage, perspective shift}, and \textit{Filter}) and the usage patterns. In addition, we incorporated quantitative survey results with qualitative data to assess how \system{} affected participants’ workload.

\subsubsection{The Usefulness of Instrumental Operations}\label{instrument_interation}

\begin{figure*}[t]
  \centering
  \begin{minipage}[t]{0.32\textwidth}
    \centering
    \includegraphics[width=\linewidth]{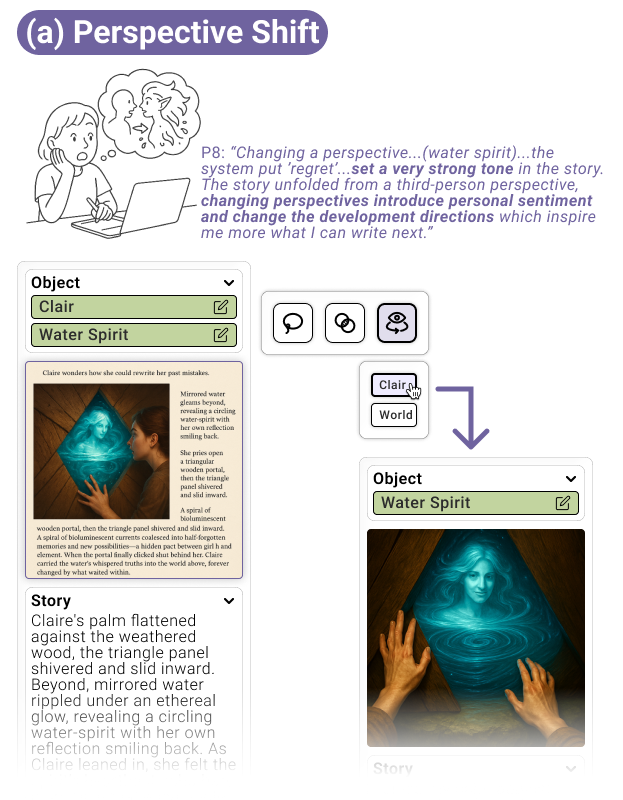}
  \end{minipage}
  \hfill
  \begin{minipage}[t]{0.32\textwidth}
    \centering
    \includegraphics[width=\linewidth]{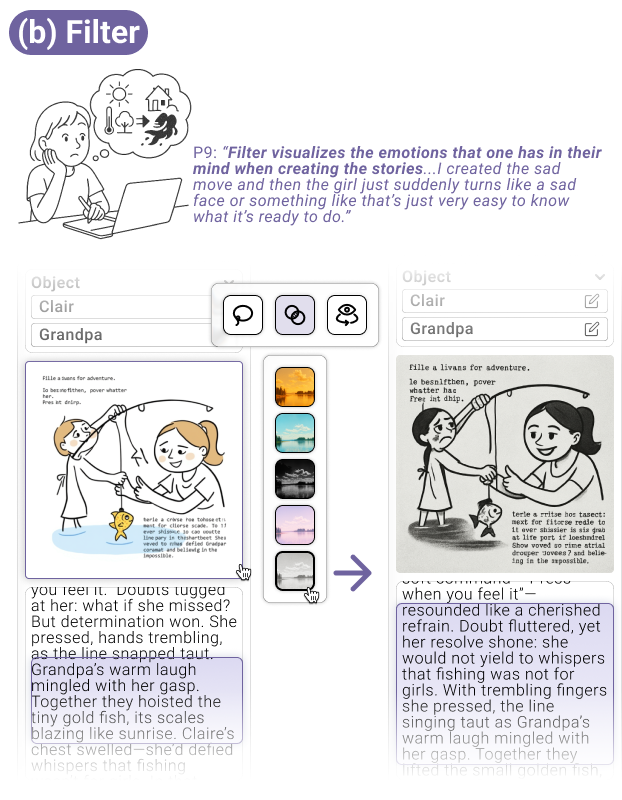}
  \end{minipage}
  \hfill
  \begin{minipage}[t]{0.32\textwidth}
    \centering
    \includegraphics[width=\linewidth]{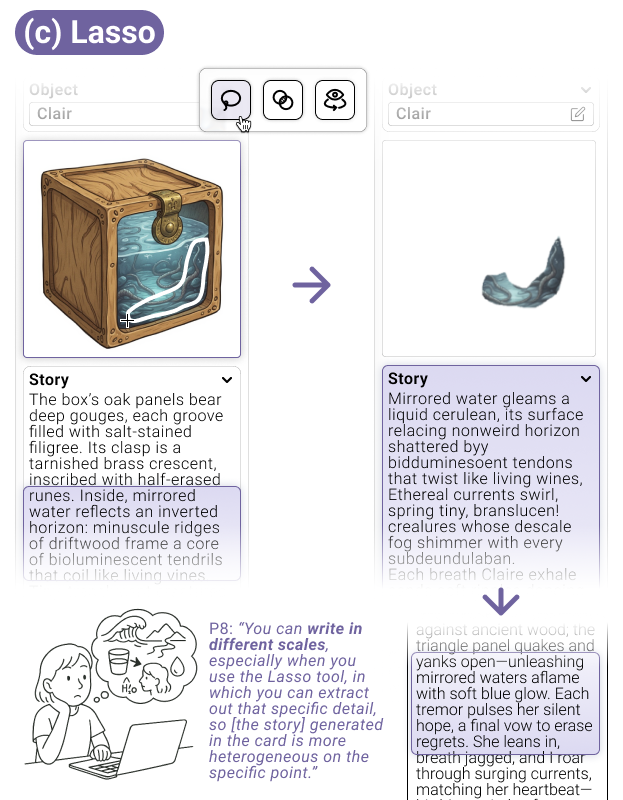}
  \end{minipage}

  \caption{
  (a) \textbf{Perspective Shift} changes image viewpoint to reframe narrative voice and redirect story development;
  (b) \textbf{Filter} synchronizes mood and style across media—applying a visual Filter also rewrites the associated text to match the intended emotional tone;
  (c) \textbf{Lasso} enables participants to operate at different narrative scales by extracting or isolating elements to steer local and global edits.
  }
  \label{fig:Instruments}
  \Description{The figure provides examples of instrumental operations in use. It shows how Perspective Shift reframes narrative voice, how Filters synchronize emotional tone across images and text, and how the Lasso enables writers to operate at different narrative scales by isolating or extracting specific elements for local or global edits.}
\end{figure*}

\paragraph{Lasso as a granularity controller for local-to-global rewriting.}
The \textit{Lasso} instrument helped participants shift between narrative scales, supporting both macro-level story planning and micro-level detail refinement. As P8 noted, \textit{``You can write in different scales, especially when you use the Lasso tool, in which you can extract out that specific detail, so [the story] generated in the card is more heterogeneous on the specific point.''} (Figure~\ref{fig:Instruments}(c)). Similarly, P9 and P11 described using the Lasso to foreground key points and sustain more focused attention during revision. \added{Together, these accounts suggest that the Lasso affords a narrative “zoom,” enabling users to isolate a local fragment for targeted rewriting while maintaining continuity with the broader narrative trajectory within the same interface.}

\paragraph{Collage Function Enables Creative Recombination.}

\begin{figure*}[htbp]
  \centering
  \includegraphics[width=\textwidth,height=0.45\textheight,keepaspectratio]{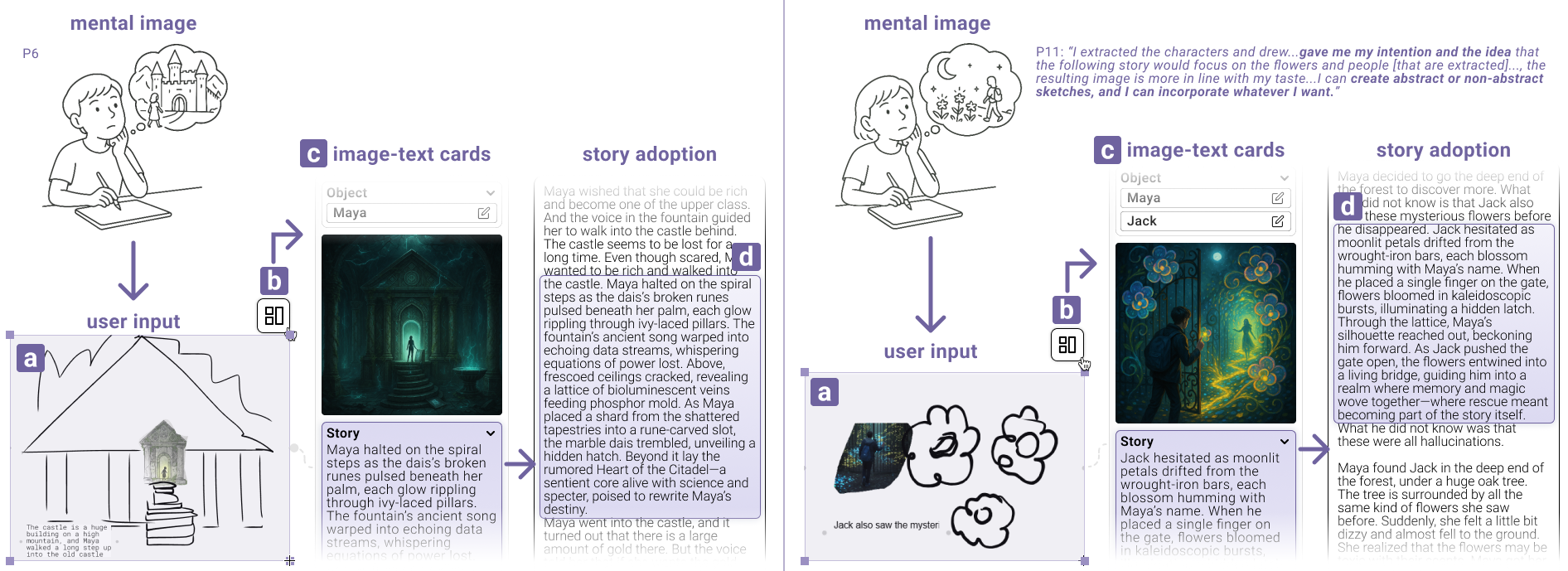}
  \caption{Collage. Participants used Collage for creative recombination, merging extracted objects/scenes (often mixing sketches with images) to specify visualization and advance story ideas; usability was positively rated.}
  \label{fig: Collage}
  \Description{The figure demonstrates the Collage operation during story writing. Participants combine extracted objects and scenes from multiple cards, often mixing sketches with images, to specify visualization and explore new narrative ideas. The example highlights creative recombination as a mechanism for advancing story development.}
\end{figure*}

Participants used Collage to merge extracted objects or scenes from different images, building connections across disparate elements.
For instance, when P6 generated a scene of Maya entering a castle, he envisioned a larger structure with taller stairs. He sketched a bigger castle and mountain while expressing his intention through text, resulting in a generated card that matched his vision and was directly incorporated into his story.
P11 articulated the creative freedom Collage technique provided: \textit{``This technique doesn't limit me; I can create abstract or non-abstract sketches, and I can incorporate whatever I want.''} (Figure~\ref{fig: Collage}). \deleted{This multimodal recomposition enabled participants to quickly express envisioned scenes (P2) and provided ``more freedom to envision and create the story'' (P10). These practices highlight that Collage is not merely a usability feature but a catalyst for multimodal recomposition, enabling participants to externalize, reconfigure, and expand their mental imagery into coherent narrative possibilities that text or images alone could not achieve.}\added{Collage thus serves not merely as a usability feature but as a catalyst for externalizing, reconfiguring, and expanding mental imagery into narrative possibilities unachievable through text or images alone.}

\paragraph{Perspective Shift as a narrative frame-shifter.}
 The \textit{Perspective Shift} alters both the image and the story perspective to provide new directions for story development. P8 described how shifting perspectives changed the story direction: \textit{``Adopting the water's viewpoint anthropomorphized the water spirit and introduced a regretful undertone that established the story's emotional framework... changed the development direction, inspiring new writing possibilities.''} P5 also experimented with this feature, incorporating a first-person voice \textit{(``I didn't expect this to be so heavy!'')} adopted from the system-generated segments into her third-person story (Figure~\ref{fig:Instruments} (a)). Perspective Shift allowed participants to flexibly reconfigure narrative viewpoint and voice, surfacing new emotional framings and redirecting story trajectories without disrupting their ongoing writing flow.

\paragraph{Filter as affective parameterization for tone alignment.}
The \textit{Filter} instrument shaped narrative emotion and tone by visually parameterizing affect. P11 noted, \textit{\deleted{``The image provides the style, which influences my story's tone and direction... Before using the Filter, I can’t determine tone from text alone---I need to choose between suspenseful or romantic expression ways. However, visual changes after applying a Filter help me decide which feeling I want my text to have.''} \added{``The image's style influences my story's tone and direction. Text alone doesn't reveal tone. The Filter's visual changes help me decide the desired feeling when I need to choose between suspenseful or romantic approaches.''}}
Similarly, P9 observed, \textit{``Filter visualizes the emotions in my mind when creating stories... I created a sad mood with Filters, and the girl suddenly turned into a sad face. It’s very easy to see what it’s doing and easy for me to describe later.''}
(Figure \ref{fig:Instruments} (b)). The immediate visual feedback aligned emotional intent with text, streamlining tone-setting decisions to evaluate the usability.

Taken together, these instrumental operations transformed localized operations into meaningful viewpoints, scales, and tones. They appear to support participants’ intended operational precision and expressiveness, while potentially reinforcing the perception–action loop during the planning and translating of story writing.

\subsubsection{Tradeoffs of Multimodal Image-Text Co-editing}

\paragraph{Increased Workload.} While image–text co-editing may support aspects of narrative coherence and expressiveness, our results suggest that it can also introduce additional workload. NASA–TLX scores showed significantly higher mental demand (NASA-TLX: $M_{\text{\system{}}}=5.16$ vs.\ $M_{\text{baseline}}=3.17$, \textit{p}=.0000; Table~\ref{table:Statistics}) and physical demand (NASA-TLX: $M_{\text{\system{}}}=4.67$ vs.\ $M_{\text{baseline}}=2.08$, \textit{p}=.0002; Table~\ref{table:Statistics}), plus moderately higher frustration (NASA-TLX: $M_{\text{\system{}}}=2.75$ vs.\ $M_{\text{baseline}}=1.75$, \textit{p}=.0204; Table~\ref{table:Statistics}). Qualitative analysis reveals that this increased workload reflected the higher cognitive and physical effort of coordinating across modalities and actively curating outputs compared with the GPT baseline \deleted{which involves entering text prompts and receiving output directly}. Part of the mental load may have stemmed from first-time use—learning new image/text operations and switching between modalities (P1, P2, P5, P9, P10). \deleted{As P2 suggested, the biggest burden is switching between tools to sketch or type; it takes time to learn and adapt, even though the functions are useful.''} Additional difficulty also arose from unfamiliarity with the canvas interface compared with the traditional GPT interface (P3, P10). \added{Below we present the tension of Image-Text co-editing based on qualitative feedback.}

\textit{The Cognitive Effort of Enhanced Sense of Agency.}
\deleted{Nine of the participants valued the ability to maintain control of the story, and the creation process gave them a stronger sense of agency (all participants except P2, P7, and P12), and participants felt that they were directing the story’s development---the final narrative emerged from their own sketches, inputs, and use of the system's instrumental operations (P4, P6, P10). } \added{Nine participants (all except P2, P7, P12) felt they have higher sense of agency, and valued maintaining story control and felt they were directing the narrative through their sketches, inputs, and use of system tools (P4, P6, P10).} However, P1, P3, P9, and P10 noted that they had to actively develop details within their own text, especially at the early stages, which could feel ``a little bit frustrating'' (P9). Unlike GPT, which could quickly produce long passages or propose questions to guide brainstorming (P10), the system requires participants to supply and elaborate on their own ideas before meaningful generation occurs, potentially leading to a higher mental workload. 
This shift demanded more cognitive and physical effort: even though sketching and annotation helped externalize mental imagery, participants noted that it felt more demanding than simply inputting and receiving GPT’s ready-made text (P1, P2, P11). Thus, a stronger sense of agency may come at the cost of a higher workload.

\textit{Validating Ideas Rather than Generating Them.}
Some participants noted that the system's strengths lay in validating or expanding existing concrete mental images rather than generating new and abstract directions (P1, P11). As P1 explained, \textit{``When I have a vague impression in my mind, I tend to generate some image-text pairs. But sometimes, once the visual appears, it fixes my imagination in a certain way in my mind, and I can no longer imagine other possibilities. In contrast, only plain text can inspire limitless imagination.''} This reveals a tension: images act as concrete anchors that aid detailed development, yet their representational specificity can induce fixation by prematurely crystallizing fuzzy concepts and narrowing exploration. \looseness=-1
Similarly, P10 noted that the baseline GPT condition was superior at breaking down initial story points and directly providing additional suggestions and directions, whereas the \system{} system primarily served to elaborate or diverge from existing visions the user already has. This suggests this multimodal approach may be valuable for participants with partially formed concepts, though potentially less helpful during the open-ended phases of ideation when abstract exploration is more important than visual specificity. \looseness=-1

\textit{Externalization Frees Cognitive Space.}
Although participants reported experiencing higher mental workload, several accounts suggest that multimodality may have supported a more efficient allocation of attention (P4, P7). 
As P7 described, \textit{``By highlighting and collaging, I externalized formed ideas into image–text pairs, clearing mental space to pre-plan the next line and concentrate on the next plot beat.''}
It indicates that the creation process has the potential to externalize fleeting ideas, preserve sensory and spatial detail, and reduce information loss when translating imagination into concrete artifacts. 
\deleted{These accounts indicate that while \system{} demands more decision-making effort, it also enables a degree of cognitive offloading that shifted attention away from low-level memory maintenance toward higher-level creative synthesis. With greater familiarity, such offloading could potentially yield efficiency benefits that offset the initial overhead.}






\subsection{Creativity Support of \system{}}
\label{6.2}

To address RQ3 and RQ4, we examine how \system{} supports intention expression through multimodal input, facilitates the ideation process through divergent exploration, as well as how image–text alignment design contributes to narrative development. We also describe how this workflow potentially enhances participants’ sense of agency and ownership.

\begin{figure*}[htbp]
  \centering
  \includegraphics[width=\textwidth,height=0.45\textheight,keepaspectratio]{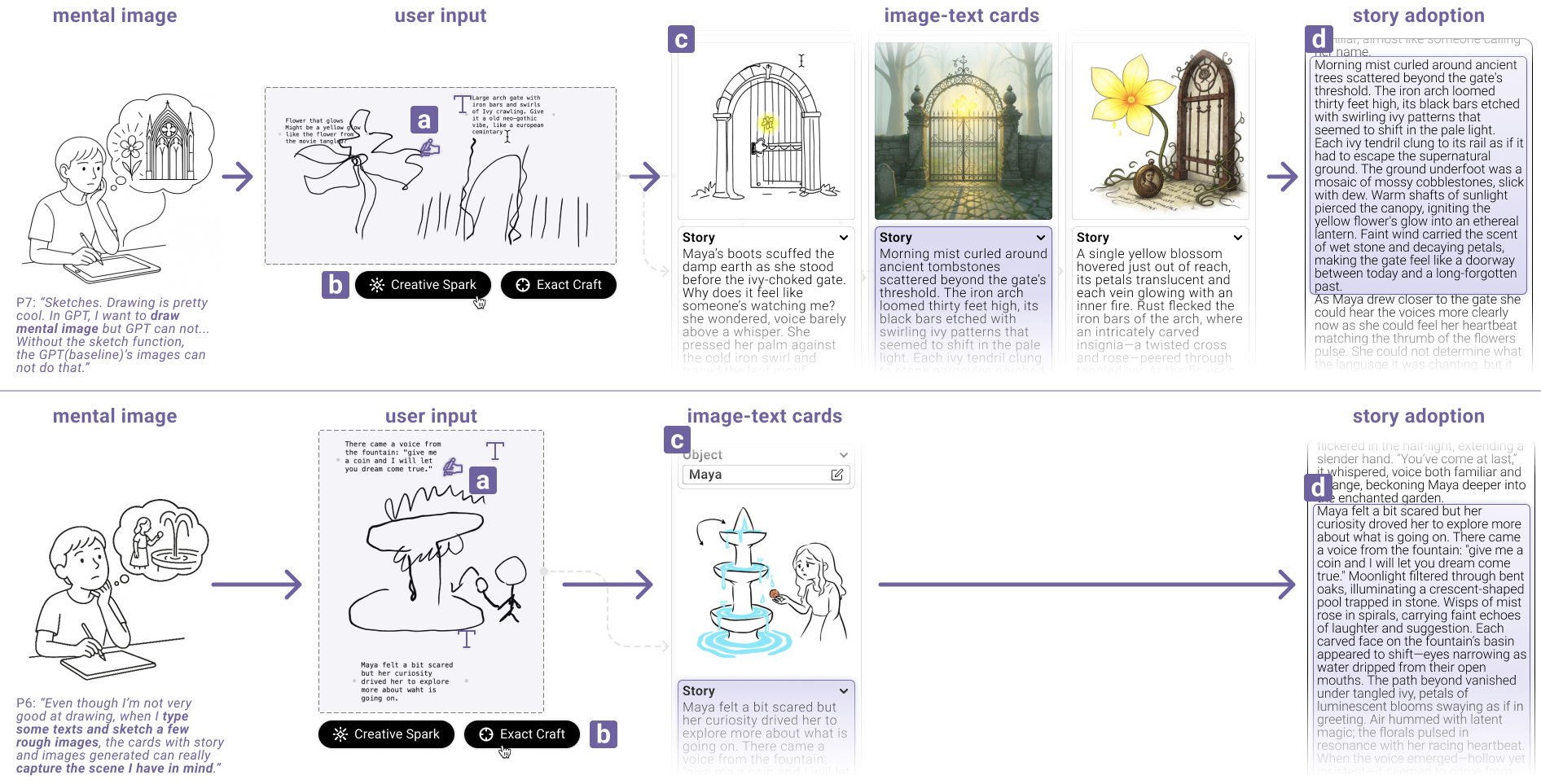}
  \caption{Multimodal Expression. Example where sketch structure and textual details jointly yield multiple relevant images; participants valued sketches for visualizing spatial layout beyond what text-only tools could provide.}
  \label{fig: Result - Multimodality}
  \Description{The figure shows an example of multimodal expression where rough sketches and textual descriptions jointly guide image generation. The resulting images reflect the spatial layout and visual structure conveyed by sketches that would be difficult to express using text alone, supporting more precise externalization of mental imagery.}
\end{figure*}

\subsubsection{Enhancing Intent Expression Through Multimodal Input.}
As shown in Appendix Figure \ref{fig:logbar} (b), text served as the primary medium for generation but was consistently supplemented with sketches and images to provide more spatial information. Eight participants (P1, P2, P3, P6, P7, P9, P10, P11) stressed that combining sketches with text and images aligned outputs more closely with their creative intentions, yielding significantly higher expressiveness ratings than the baseline ($M_{\text{Vistoria}} = 6.08$ vs.\ $M_{\text{Baseline}} = 4.33$, \textit{p} = .023; Table \ref{table:Statistics}).

P6 captured this benefit: \textit{``Even though I'm not very good at drawing, when I type some texts and sketch a few rough images, the cards with story and images generated can really capture the scene I have in mind.''}.  P7 illustrated this with a concrete case: she sketched a rough flower and archway, then added text specifying a glowing flower and a Gothic gate. The system fused the spatial layout from the sketch with textual details to generate multiple fitting images. She highlighted the unique value of sketching: \textit{``Drawing is pretty cool. In GPT, I want to draw a mental image, but GPT cannot... the geometry of GPT-generated image is always different from what’s in my head.''} (Figure~\ref{fig: Result - Multimodality}).
This suggests that multimodal expression enabled participants to externalize their mental imagery and refine it into concrete, shareable representations, bridging the gap between vague internal visions and precise outputs.

\subsubsection{Bottom-up Creation to Support Divergent Exploration}
As shown in Appendix Figure~\ref{Iteration_Process} and Appendix Table~\ref{tab:descriptive}, participants demonstrated greater breadth and depth of exploration using \system{} and produced visibly more divergent narrative structures compared to the linear outputs of the GPT baseline. The \system{} canvas functioned as an exploratory space where participants pursued multiple storylines in parallel without linear constraints. 

Rather than committing immediately to single narratives, participants typically generated multiple alternatives in early phases. \deleted{, positioning the system as an expressive medium rather than merely a text generator (P7, P4, P5, P8). This exploratory approach helped participants avoid early fixation and sustained creative engagement} As P8 observed: \textit{``GPT workflow is more streamlined... top-down. Using the system feels more bottom-up. You are open to possibilities, and then you choose one way to go deep, so there's not a finite result and more possibilities being explored.''} 

The exploratory nature of \system{} was also described as playful and enjoyable (P1, P7). As P1 reflected: \textit{“Using the tool feels more like doing Collage or drawing on a whiteboard or a large sheet of paper, where you can do almost anything—it's very free and interesting.”} This experiential quality aligns with the higher immersion reported for \system{} compared to the baseline (CSI: $M_{\text{Vistoria}} = 4.92$ vs.\ $M_{\text{Baseline}} = 2.75$, \textit{p} = .0006; Table~\ref{table:Statistics}).

\begin{figure*}[t]
    \centering
    \includegraphics[width=\textwidth]{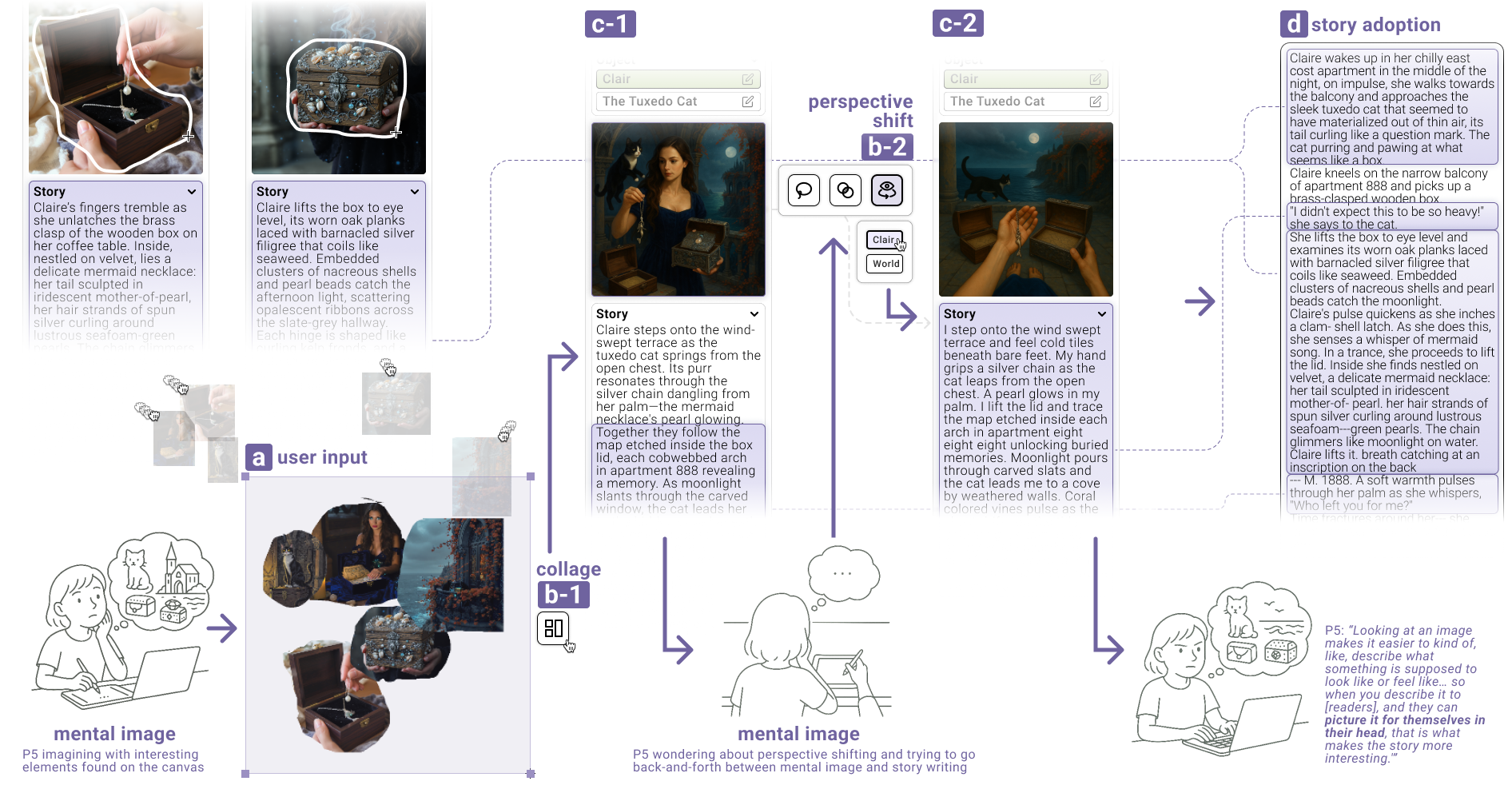}
    \caption{Images act as cognitive scaffolds, helping participants describe unfamiliar actions or contexts more concretely. P5 constructed her final writing by referring to both images and adopting and editing related text.}
    \label{Multi}
    \Description{The figure shows an example canvas during early exploration, where multiple image–text cards, sketches, and annotations are distributed spatially. Different narrative directions are pursued in parallel, illustrating non-linear exploration before committing to a single storyline.}
\end{figure*}


\subsubsection{Leveraging Image-Text Alignment for Rich Descriptions and Story Progression}
When text and images appear together, the two modalities reinforce one another, enabling unfamiliar or imagined elements to be both visualized and verbalized. This cross-modal grounding supports the production of more detailed and enriched descriptions. 

Visual references, in particular, helped participants imagine actions or settings beyond their lived experience. As P5 noted, \textit{``Looking at an image makes it easier to kind of, like, describe what something is supposed to look like or feel like... so when you describe it to [readers], and they can picture it for themselves in their head, that is what makes the story more interesting.''} From observation, P5 closely described the scene in her imagination with the cliff, the necklace, and the cat in the image in her writing and directly adopted some text generated, integrating them into her final story (Figure~\ref{Multi}).

Using the generated text and image also helped produce more concrete, detailed narratives. For example, P1 used visual cues from an image showing Claire touching a letter and adopted the descriptions in the text, such as \textit{``Claire steadies the box on her lap''} and created the narrative of \textit{``She runs her fingers over the letters, heartbeats echoing in her ears.''} to form his final story. Here, the text description with visualization allow participant to capture more dynamic, sensory-rich narrative moments.

\deleted{Furthermore, unlike traditional workflows where participants must read through text to review their progress, these image-text pairs also facilitate easier tracking of story development and reduce idea drift or loss (P12, P11, P2). P12 noted that visuals alongside text helped maintain the mood and recall earlier ideas. P2 further described how the visual sequence supported planning:
Using the system, we started with an initial image and then came up with another image… seeing the visual sequence made it easy to trace the development of the story. If you get too far down that chain and don't like it, you can just delete that node and go in a different direction. I liked visually being able to see the progress from generation to generation.'' Having images presented alongside text also helped participants manage the development of their narratives and better understand where the story should go next.
}

\subsubsection{Preserving Sense of Agency and Ownership.}
Unlike the baseline, where participants often felt like they were editors of GPT-generated text, \system{} supported exploratory editing while preserving a sense of agency and ownership. Participants felt that \system{} ``having more sense of mastery over the content,'' in contrast to the baseline (P4, P6, P9). P7 felt that when using \system{}, she was actively cutting, combining, filtering, and directing the story’s trajectory, whereas with GPT, she was mostly receiving and editing what the model produced. P9 explicitly emphasized a heightened sense of agency, noting: \textit{``This tool is more `me'... I control characters and plots.''}

Dissatisfied with the baseline condition, in which GPT produced most of the content and left participants primarily in the role of adopting it (P4, P6, P7), P10 also characterized \system{} as a supportive co-pilot rather than a substitute for their own work. Quantitative results are consistent with these perceptions: participants rated \system{} as providing a stronger collaborative experience than the baseline (CSI: $M_{\text{Vistoria}} = 5.50$ vs.\ $M_{\text{Baseline}} = 4.58$, \textit{p} = .0418; Table~\ref{table:Statistics}). \deleted{which participants interpreted as a `co-pilot' relationship that preserved their sense of agency. }

This preserved sense of agency also led to the preserved sense of ownership. GPT outputs in the baseline condition were repeatedly described as ``surface--level'' (P2, P5, P7) and as ``someone else's work'' (P3), whereas most participants reported a stronger sense of ownership with \system{} (P1, P2, P3, P4, P5, P6, P8, P9, P10, P11). As P5 explained: \textit{``When using \system{}, every idea originated from my own imagination, and the final story was formed by manipulating and combining these different self-generated ideas. This gave me a strong sense that the story was truly my own creation.''}

Together, these findings suggest a shift from passively adopting model suggestions to actively creating and curating one’s own generative outputs.

%% file: sections/06-Discussion.tex
\subsection{Multimodal Instrumental Interaction}
Instrumental Interaction~\cite{beaudouin2000instrumental} conceptualizes instruments as mediators that translate writers' actions into operations on domain objects. We operationalize this principle by reifying a set of multimodal instrumental operations (Collage, Lasso, Filter, and Perspective Shift) that simultaneously act upon both text and image narrative materials. Rather than treating images and text as separate interface elements, these instruments serve as unified interactional units that enable writers to zoom between narrative scales, reorganize multimodal story fragments, and explore divergent directions within a shared representational space to edit image and text simultaneously.

From the perspective of \textit{Designing Interaction, not Interfaces}~\cite{10.1145/989863.989865}, our design moves away from adding more interface widgets and focuses instead on shaping the quality of writers’ ongoing activity. Beaudouin-Lafon argues that transformative interfaces must shift attention from surface-level user-interface (UI) components toward the underlying interactional structures that support creative work~\cite{10.1145/989863.989865}. Following this perspective and Dual Coding Theory's suggestions for verbal and nonverbal perception alignment~\cite{clark1991dual}, Vistoria’s design prioritizes fluid transitions between modalities, persistent manipulable artifacts, and an iterative loop in which narrative ideas and multimodal materials co-evolve. This interaction-centered framing explains why writers perceived the system as increasing expressiveness.

Although participants generally found the instruments effective, they also reported a substantial learning curve. For novice system writers, the system imposed considerable mental and physical workload. These insights suggest future design opportunities to lower workload and adapt to writers’ evolving needs. In the early stages of system use, writers could express their intentions through natural language, allowing the system to suggest the most appropriate functions on their behalf~\cite{10.1145/3715336.3735766}. As writers become more familiar with the system, they can choose functions by themselves and even define customized functions that better fit their evolving needs. This approach aligns with the emphasis on supporting diverse, situated practices rather than enforcing a fixed interface vocabulary~\cite{10.1145/989863.989865}. Together, these directions point toward a design space where multimodal creativity systems integrate explicit instruments with adaptive, activity-centered interaction models to better support real-world writing workflows.



\subsection{Designing \added{Dynamic} Mixed-Initiative Multimodal Workflows}
In the traditional turn-based GPT workflow, writers often occupy a relatively passive or evaluative role: they receive model output and act primarily as examiners who check, accept, or correct the result~\cite{kwan2024mt}. In \system{}, writers actively manipulate elements on the surface, decide which multimodal materials to combine, and select which operations to apply. From the perspective of the participants, \system{} is not experienced as a detached “answer engine,” but as a co-pilot collaborator. Writers perceive themselves as those who decide how to create, what to keep, and which tools to invoke, preserving the sense of agency and ownership. 

However, this also brings cognitive effort. Precisely because the workflow is instrument and manipulation-driven, it demands that writers have a clearer sense of what they need, or at least which direction they wish to explore. When writers do not yet know what they want, the system requires them to specify intentions and choose operations that tend to lead to higher cognitive load. \added{This reflects an inherent tension: high functional agency (agency is structurally distributed in the system) requires clear intentionality, but early-stage ideation is often characterized by ambiguity and exploration~\cite{Rafner01122025}}. In contrast, a text-only GPT chat enables the rapid generation of a large amount of content, allowing for subsequent refinement through iterative prompting and selection to achieve a specific focus. \deleted{Several participants, therefore, viewed our system as especially suitable when they already had some ideas or a tentative direction, rather than when they were starting from a completely blank slate.} \added{This suggests that systems should adapt their agency distribution across creative phases as agency and ownership are not static end-states but fluctuate across the co-creation trajectory~\cite{Rafner01122025}. In early divergent exploration, lower functional agency may paradoxically support higher felt agency (e.g., through control, decision-making, or ownership) by reducing decision fatigue; in later convergent refinement, higher functional agency (granular control) aligns with writers' desire to shape details~\cite{Rafner01122025,fu2024being}.}

\deleted{This inspires us to design a mixed-initiative paradigm to enable smooth transitions between model-led and user-led modes to accommodate different control~\cite{10.1145/302979.303030}. When writers lack a clear direction, the system should allow temporary shifts toward more GPT-like exploration, for example, generating diverse suggestions or story seeds that can then be brought back onto the canvas for instrumental refinement. It could also offer low-commitment ways to switch modalities; when intentions are clearer, it should allow more user autonomy, like fine-grained instrumental operations for precise control. Furthermore, supporting mixed-initiative involves more than simply adding a model-led mode; it also requires careful design of how and when transitions between user-led and model-led states occur. This suggests designing meta-instruments that regulate the division of labor between user and model. For example, when writers use the Collage function, they can request the model to recommend collage direction, and useful elements potentially can be involved based on the existing writer's intention to realize a true collaborative activation of ideas and shared cognition between the writer and the model. In this framing, mixed-initiative becomes an additional layer of instrumental interaction that allows users to explicitly shape who drives which parts of the creative process.}

\added{On the other hand, ownership emerges from users' active control over outputs and their ability to modify content~\cite{10.1145/3637875}. Critically, this control includes not only refinement flexibility but also rejection. Our results show that participants sometimes think image output ``fixed imagination in a certain way,'' which may threaten to constrain their creative vision. Previous work indicates that when AI outputs deviate from expectations, users experience diminished control and ownership uncertainty~\cite{10.1145/3637875}, and encouraging system design to prompt them to deploy adaptive strategies to restore agency. In that case, future designs could operationalize this through a function that, when invoked, asks ``What specifically do you want to avoid?'' and generates alternatives that explicitly diverge from the rejected output. This would transform counter-inspiration~\cite{Rafner01122025} from an ad-hoc user strategy into operations that reframe dissatisfaction from a null action into a mechanism for articulating and preserving authorial boundaries and ownership.}

\subsection{Limitation and Future Work}
While our study provides valuable insights into multimodal story writing, several limitations constrain the generalizability and scope of our findings.

\textit{Task Scope and Short-Term Focus.} The 300-to-500-word story task, while manageable for controlled evaluation, does not reflect the demands of long-form fictional writing, where authors build sustained voices, complex arcs, and intricate structures~\cite{chen_once_2025}. We also did not have an opportunity to study \system{}'s suitability for extended projects, iterative revisions, or complex narratives, leaving open questions on the consistency in long works, scalability with larger volumes, or risks of over-reliance on LLM over time of use. Moreover, evaluation relied primarily on self-reports of creativity and user experience; we did not include objective measures of story quality, originality, or literary merit. 

In the future, we plan to conduct field studies that deploy this system with writers of varying expertise levels in their authentic creative contexts, observing how they integrate the tool into real writing projects over extended periods. Such longitudinal research would assess the ecological validity of \system{} and provide insights into how writers adapt \system{} for planning and translating across longer creative cycles~\cite{10.1145/3746058.3758469}. We anticipate that writers might spontaneously capture inspirational moments from daily life, potentially increasing their reliance on clustering functionality as they generate more dispersed content fragments that require organization. These naturalistic studies would provide crucial insights into the tool's role in sustained creative practice, revealing usage patterns, adaptation strategies, and long-term impacts on writers' creative processes that controlled laboratory settings cannot capture.

\textit{Participant Sample.} Our study involved only 12 participants, while this is typical for similar lab usability studies, a larger group could provide stronger statistical power, reveal more varied interaction patterns, and allow comparisons across subgroups. Furthermore, the group of participants has limited cultural and age diversity, which could have narrowed the range of narrative traditions, writing styles, and storytelling approaches represented. Future research should address these limitations by recruiting a larger and more diverse set of participants, including writers of various ages and individuals from diverse cultural backgrounds, to more fully evaluate the applicability and generalizability of the system.

\textit{Construct Validity and Measurement Limitations.} Although we discuss constructs such as creativity, sense of ownership, and agency, these observations arise primarily from qualitative reports. Our study does not include construct-grounded measurements or comparative baselines for these phenomena. Accordingly, the interpretations should be viewed as exploratory insights rather than empirically validated effects. Future work will incorporate construct-aligned measures and validated scales such as the Mixed-Initiative CSI~\cite{10.1145/3581641.3584095}, \added{or using frameworks situated within human–AI co-creativity like COFI~\cite{10.1145/3519026} and CCDF~\cite{10.1145/3698061.3726934} to quantify co-creativity interaction dynamics or better derive Human-AI co-creativity insights~\cite{10.1145/3769072}.}

\added{Furthermore, while we assessed participants' AI usage frequency (e.g., daily ChatGPT usage), we did not measure their collaborative AI literacy or metacognitive capabilities when working with AI systems. 
Future research should incorporate validated scales ~\cite{Sidra19082025} to investigate how 
collaborative AI literacy influences the effectiveness of multimodal co-editing tools.}

\textit{Multimodality Scope.} Our work focuses on multimodal support through text and images, but does not include other modalities. Prior research in creative writing suggests that audio can also serve as a useful medium~\cite{10.1145/3511599}, especially through nonverbal sounds and ambient effects that help shape mood and atmosphere. In future work, we plan to explore the addition of audio cues to the writing process. Such sound elements may support writers in building a stronger vibe, enhancing scene-setting, and offering an additional channel for creative inspiration~\cite{10.1145/3519026}.

%% file: sections/07-Conclusion.tex
This paper presents \system{}, a multimodal image–text co-editing system that supports fictional story writing by tightly integrating image and text representations. Grounded in the WoZ co-design study, Vistoria introduces a unified set of instrumental operations (Lasso, Collage, Perspective Shift, and Filter) that reify writers’ intentions and enable synchronized manipulation across modalities. Through a controlled user study, we demonstrate that multimodal co-editing enhances expressiveness, immersion, and exploratory ideation. Although this multimodal workflow increases cognitive demand, participants reported preserved senses of agency and ownership, treating the system as a creative partner rather than a generative tool. We hope \system{} highlights the opportunities for designing future writing systems that embrace multimodality as a core mechanism for ideation and narrative development.

%% file: sections/Appendix.tex
\subsection{User Behavior in the WoZ co-design study}
\label{tab:formative_grouped}

\begin{figure}[H]
    \centering
    \includegraphics[width=1\linewidth]{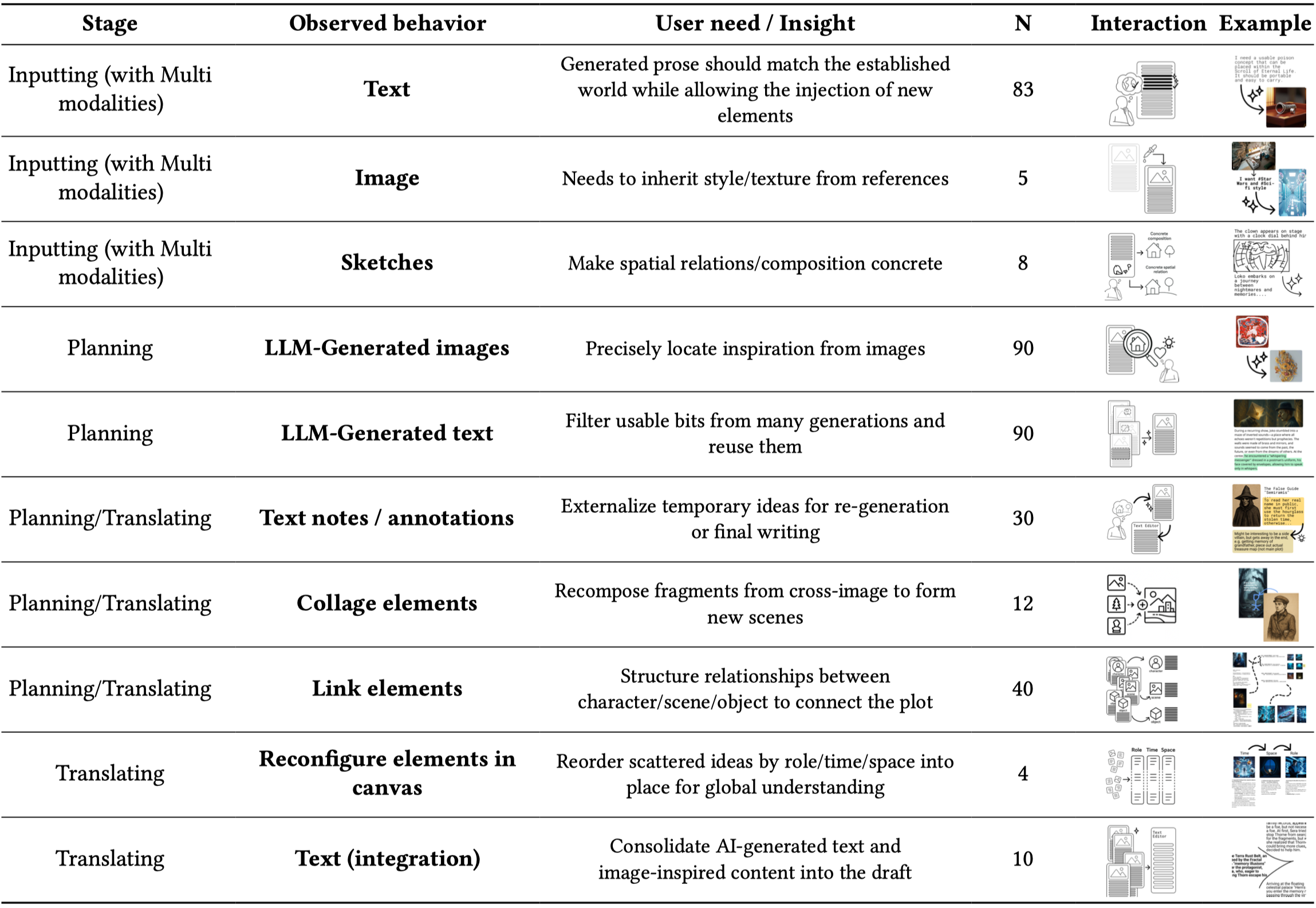}
    \caption{Results of user strategies for manipulating multimodal elements in the WoZ co-design study.}
    \Description{Results of user strategies for manipulating multimodal elements in the WoZ co-design study.}
\end{figure}

\clearpage
\subsection{Behavioral interaction data gathered from participants in the user study}

\begin{figure}[H]
  \centering
  \includegraphics[width=0.8\linewidth]{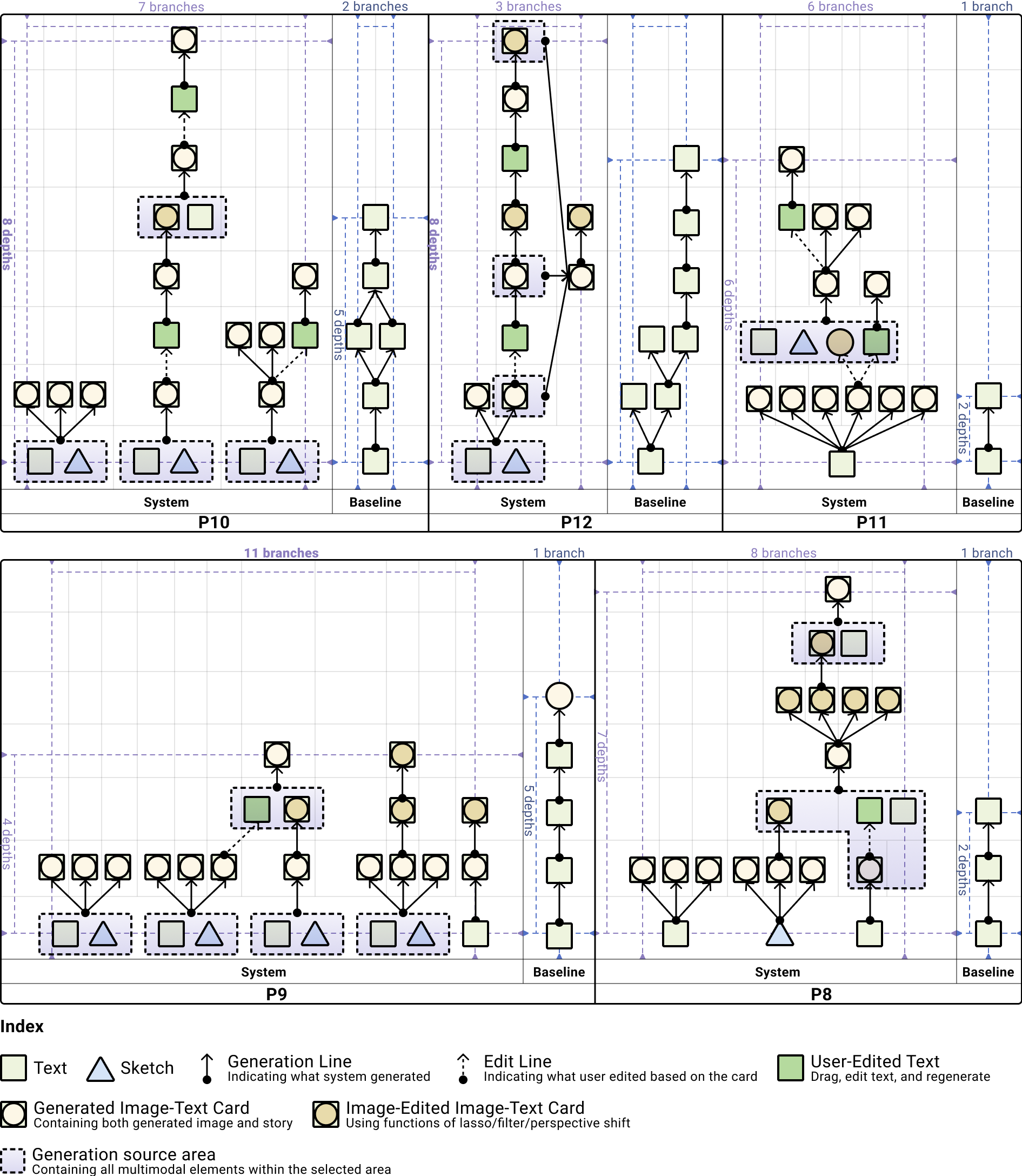} 
  \caption{Behavioral diagram contrasting Vistoria vs. baseline: participants cycle through multimodal generation, collage/recombination, and coupled image–text editing before collecting highlights for integration. Data from P8-P12 show that, compared with baseline, participants using Vistoria explored more directions, with greater divergence (Branches) within each direction. Participants also tended to pursue deeper exploration within specific directions when using Vistoria.}
  \label{Iteration_Process}
  \Description{The figure compares story development outcomes between the baseline text-only condition and Vistoria. It visualizes differences in narrative branching and structural diversity, showing how Vistoria supports parallel exploration and alternative story trajectories.}
\end{figure}

\clearpage
We examined the sequences of function use and compared exploration patterns across the baseline and our condition. To characterize participants’ divergent–convergent behaviors during the creative process, for each task, we reconstructed exploration structures by defining Directions as top-level trajectories toward a goal, Branches as the diversity of possibilities generated within a direction, and Depth as the mean number of iterative steps within each branch to compare the exploration across the baseline and our condition. 

\begin{table}[H]
\centering
\small
\caption{Descriptive statistics (mean $\pm$ SD) for \system{} vs. Baseline. When using \system{}, participants exhibited broader exploration; at the same time, as shown in Figure~\ref{fig:logbar}, they also tended to pursue individual directions with greater depth. Specifically, Directions denote the number of distinct aspects or dimensions explored when co-creation with \system{} or baseline. Branches represent the diversity of possibilities generated within a given direction. Depth indicates the mean number of iterative steps within each branch. }
\label{tab:descriptive}
\setlength{\tabcolsep}{8pt}
\renewcommand{\arraystretch}{1.15}
\begin{tabular}{lcc}
\toprule
& \system{} & Baseline \\
\midrule
Mean \# of directions   & 6.92 $\pm$ 2.81 & 1.42 $\pm$ 1.08 \\
Mean \# of branches     & 3.00 $\pm$ 1.35 & 1.92 $\pm$ 0.67 \\
Mean depth              & 1.70 $\pm$ 1.18 & 2.00 $\pm$ 1.22 \\
\bottomrule
\end{tabular}
\vspace{10pt}

\end{table}

\begin{figure}[t]
    \centering
    \includegraphics[width=1\linewidth]{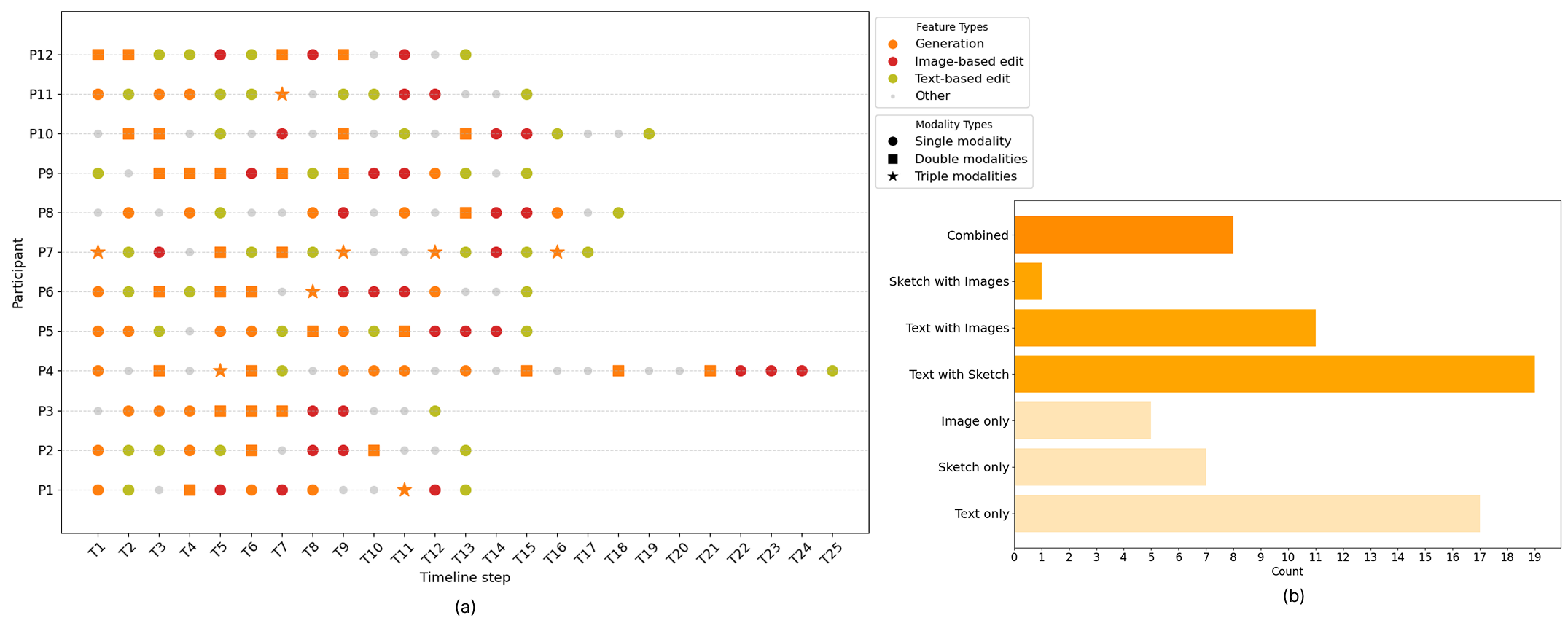}
    \caption{Interaction records of all participants. The creative workflow begins with multimodal generation—primarily text, complemented by sketches or images to express intentions, followed by refinement and iteration using visual instruments, during which textual descriptions are continuously revised in parallel. Definition of specific behaviors: Generation includes multimodal creation of new cards using Creative Spark, Exact Craft, or Collage; image-based editing refers to operations such as Lasso, perspective shift, and filters; text-based editing covers modifications of generated story segments on the canvas as well as edits made in the text editor; other operations include updates through highlight, cluster, and upgrading global settings.}
    \label{fig:logbar}
    \Description{The figure illustrates multimodal intent expression using text, sketches, and images together. User sketches provide spatial structure, while textual annotations specify narrative details. The system integrates these inputs to generate aligned image–text cards reflecting the intended scene.}
\end{figure}

\clearpage
\subsection{Writing Topics Used in the User Study}\label{Topic}
 The two writing topic prompts used in the user study were: 
 
 Topic 1---\textit{Claire steps outside her apartment and finds a small wooden box on her doorstep. The box is secured with an old brass clasp and feels unexpectedly heavy when she lifts it. Its surface is scratched, as if it has been handled many times, and faint traces of dried sea salt cling to the edges.} 
 
 Topic 2---\textit{During her morning jog through the park, Maya discovers an ornate iron gate hidden behind overgrown ivy. Through the bars, she can see a path lined with luminescent flowers that pulse gently like soft heartbeats. The air carries faint whispers in a language that sounds hauntingly familiar, almost like someone calling her name.}

\subsection{Filters}\label{fliter}
The following are the types of filter supported by the \textit{filter} instrument, showing how different types of filters are applied to image styles and mapped to text tone or emotion.

\begin{table}[h]
\centering
\caption{Filter types with corresponding image and text effects.}
\label{tab:filters}
\begin{tabular}{p{2.2cm} p{6cm} p{6cm}}
\toprule
\textbf{Filter} & \textbf{Image Effect} & \textbf{Text Effect} \\
\midrule
\textbf{Warm} & Warm tones (gold, amber, red, orange, yellow), high exposure, strong contrast $\rightarrow$ evoke happiness, comfort, nostalgia & Emphasize positivity, vitality, intimacy \\
\textbf{Calm} & Cool tones (blue, green, purple) with balanced or lower saturation $\rightarrow$ convey calmness, wisdom, introspection & Reflect contemplative and stable moods \\
\textbf{Dramatic} & Deep blacks, sharp whites, directional lighting $\rightarrow$ create intensity, mystery, urgency & Heightens stakes and emotional tension \\
\textbf{Dreamy} & Soft tones, lowered contrast, diffuse focus $\rightarrow$ suggest melancholy, intimacy, ethereality & Support subtle, nostalgic, introspective narration \\
\textbf{Monochrome} & Removal of color, emphasis on light, shadow, texture $\rightarrow$ evoke nostalgia, timelessness, artistry & Adopt reflective and universal tone \\
\bottomrule
\end{tabular}
\end{table}

\clearpage
\subsection{LLM Prompts}\label{Prompts}

\begin{center}
    \includegraphics[width=1\linewidth]{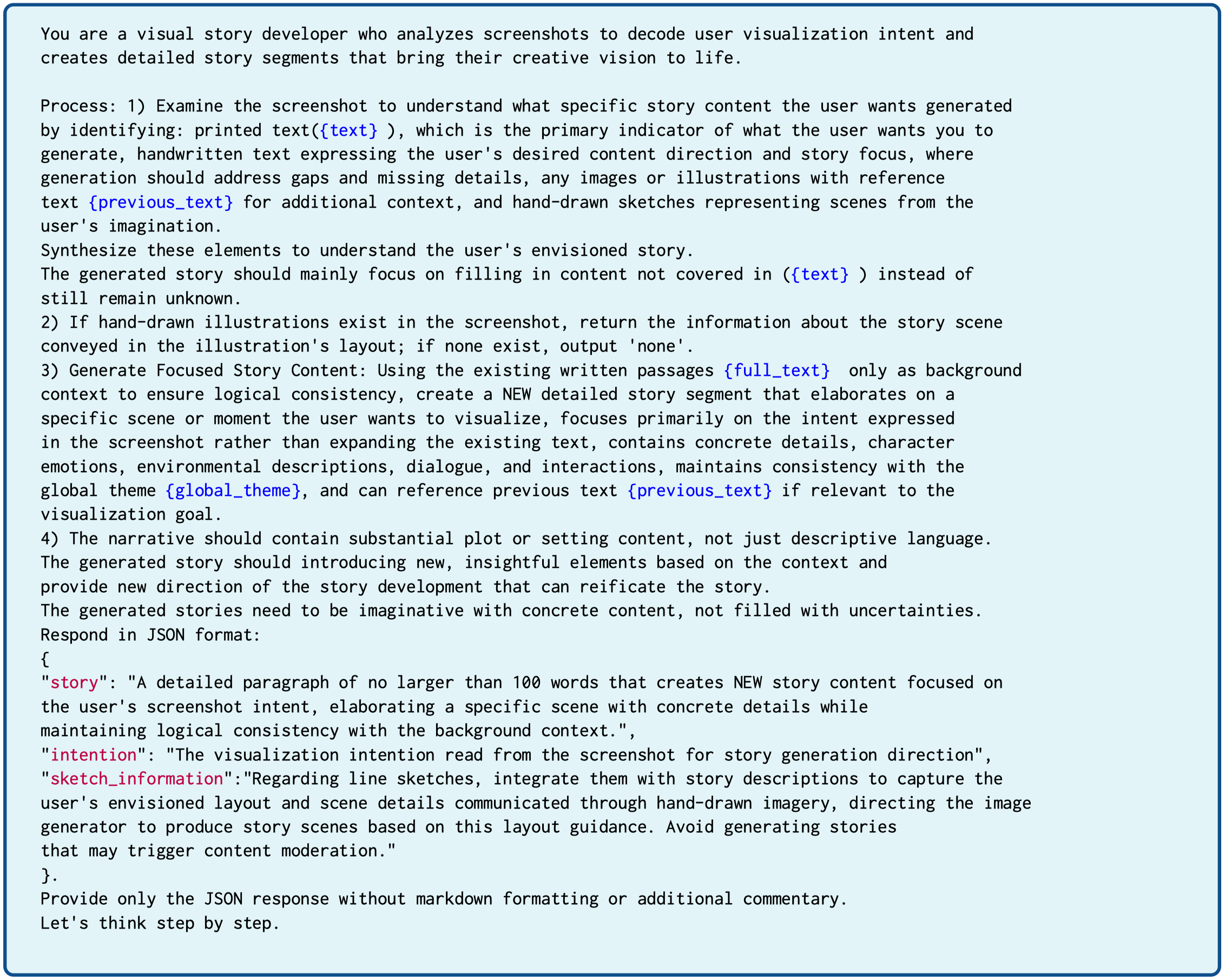}
\end{center}

\newpage
\subsection{Codebook} \label{codebook}
\added{\label{appendix:codebook}}

\begin{figure}[h]
    \centering
    \small
    \renewcommand{\arraystretch}{1.1}
    \begin{tabular}{p{\linewidth}}
        \textbf{Theme 1: Instrumental Interaction} \\
            \quad $\vdash$ \textbf{instrumental operations} \\
                \quad \quad $\vdash$ granularity control (using Lasso for detail extraction) \\
                \quad \quad $\vdash$ multimodal recombination (using Collage) \\
                \quad \quad $\vdash$ affective alignment (using Filters for tone) \\
                \quad \quad $\vdash$ perspective shift (viewpoint transformation)\\
                \quad \quad $\vdash$ open new narrative direction\\
        \textbf{Theme 2: Cognitive Process} \\
            \quad $\vdash$ \textbf{externalization \& traceability} \\
                \quad \quad $\vdash$ visual history / story evolution \\
                \quad \quad $\vdash$ spatial organization (grouping) \\
                \quad \quad $\vdash$ visual checkpoints \\
            \quad $\vdash$ \textbf{cognitive offloading} \\
                \quad \quad $\vdash$ offloading working memory \\
            \quad $\vdash$ \textbf{higher mental demand} \\
                \quad \quad $\vdash$ not familiar with operations \\
                \quad \quad $\vdash$ substantial learning effort\\
        \textbf{Theme 3: Creative Support through Multimodality} \\
            \quad $\vdash$ \textbf{bottom-up workflow} \\
                \quad \quad $\vdash$ open-ended exploration \\
                \quad \quad $\vdash$ branching storylines \\
                \quad \quad $\vdash$ comparing modalities \\
                \quad \quad $\vdash$ divergent exploration \\
            \quad $\vdash$ \textbf{inspiration} \\
                \quad \quad $\vdash$ serendipitous discovery (randomness as value) \\
                \quad \quad $\vdash$ perspective transformation\\
                \quad \quad $\vdash$ memory triggers \\
            \quad $\vdash$ \textbf{visual–text interaction} \\
                \quad \quad $\vdash$ more vivid detailed description \\
                \quad \quad $\vdash$ sense of immersion\\
        \textbf{Theme 4: Ownership and Agency} \\
            \quad $\vdash$ \textbf{control \& ownership} \\
                \quad \quad $\vdash$ resisting AI takeover \\
                \quad \quad $\vdash$ active curation \\
                \quad \quad $\vdash$ personal style alignment \\
            \quad $\vdash$ \textbf{metaphors of use} \\
                \quad \quad $\vdash$ companion / sketchbook metaphor \\
                \quad \quad $\vdash$ co-pilot\\
                \quad \quad $\vdash$ free-exploration\\
    \end{tabular}
    \caption{The final coding tree. Main themes are marked in bold; sub-codes represent specific strategies and behaviors observed in the study.}
    \label{fig:codebook_tree}
    \Description{The codebook tree of qualitative analysis}
\end{figure}

%% file: references.bib
@misc{davis2025aidrawingpartnercocreative,
      title={AI Drawing Partner: Co-Creative Drawing Agent and Research Platform to Model Co-Creation}, 
      author={Nicholas Davis and Janet Rafner},
      year={2025},
      eprint={2501.06607},
      archivePrefix={arXiv},
      primaryClass={cs.HC},
      url={https://arxiv.org/abs/2501.06607}, 
}

@article{10.1145/3519026,
author = {Rezwana, Jeba and Maher, Mary Lou},
title = {Designing Creative AI Partners with COFI: A Framework for Modeling Interaction in Human-AI Co-Creative Systems},
year = {2023},
issue_date = {October 2023},
publisher = {Association for Computing Machinery},
address = {New York, NY, USA},
volume = {30},
number = {5},
issn = {1073-0516},
url = {https://doi.org/10.1145/3519026},
doi = {10.1145/3519026},
journal = {ACM Trans. Comput.-Hum. Interact.},
month = sep,
articleno = {67},
numpages = {28}
}

@article{10.1145/3769072,
author = {Rezwana, Jeba and Maher, Mary Lou},
title = {An Exploration of Mental Models of AI in Human–AI Co-Creativity: A Framework and Insights},
year = {2025},
issue_date = {December 2025},
publisher = {Association for Computing Machinery},
address = {New York, NY, USA},
volume = {15},
number = {4},
issn = {2160-6455},
url = {https://doi.org/10.1145/3769072},
doi = {10.1145/3769072},
journal = {ACM Trans. Interact. Intell. Syst.},
month = dec,
articleno = {24},
numpages = {26}
}

@inproceedings{10.1145/3698061.3726934,
author = {Davis, Nicholas and Sherson, Jacob and Rafner, Janet},
title = {The Co-Creative Design Framework for Hybrid Intelligence},
year = {2025},
isbn = {9798400712890},
publisher = {Association for Computing Machinery},
address = {New York, NY, USA},
url = {https://doi.org/10.1145/3698061.3726934},
doi = {10.1145/3698061.3726934},
booktitle = {Proceedings of the 2025 Conference on Creativity and Cognition},
pages = {560–572},
numpages = {13},
location = {
},
series = {C\&C '25}
}

@article{gardenfors1996mental,
  title={Mental representation, conceptual spaces and metaphors},
  author={G{\"a}rdenfors, Peter},
  journal={Synthese},
  volume={106},
  number={1},
  pages={21--47},
  year={1996},
  publisher={Springer},
  doi= {10.1007/BF00172805}
}

@article{sanders2008co,
  title={Co-creation and the new landscapes of design},
  author={Sanders, Elizabeth B-N and Stappers, Pieter Jan},
  journal={Co-design},
  volume={4},
  number={1},
  pages={5--18},
  year={2008},
  publisher={Taylor \& Francis}
}

@article{vaajakallio2014design,
  title={Design games in codesign: as a tool, a mindset and a structure},
  author={Vaajakallio, Kirsikka and Mattelm{\"a}ki, Tuuli},
  journal={CoDesign},
  volume={10},
  number={1},
  pages={63--77},
  year={2014},
  publisher={Taylor \& Francis}
}

@inproceedings{qin2024charactermeet,
author = {Qin, Hua Xuan and Jin, Shan and Gao, Ze and Fan, Mingming and Hui, Pan},
title = {CharacterMeet: Supporting Creative Writers' Entire Story Character Construction Processes Through Conversation with LLM-Powered Chatbot Avatars},
year = {2024},
isbn = {9798400703300},
publisher = {Association for Computing Machinery},
address = {New York, NY, USA},
url = {https://doi.org/10.1145/3613904.3642105},
doi = {10.1145/3613904.3642105},
booktitle = {Proceedings of the 2024 CHI Conference on Human Factors in Computing Systems},
articleno = {1051},
numpages = {19},
location = {Honolulu, HI, USA},
series = {CHI '24}
}

@inproceedings{chungTaleBrushSketchingStories2022,
author = {Chung, John Joon Young and Kim, Wooseok and Yoo, Kang Min and Lee, Hwaran and Adar, Eytan and Chang, Minsuk},
title = {TaleBrush: Sketching Stories with Generative Pretrained Language Models},
year = {2022},
isbn = {9781450391573},
publisher = {Association for Computing Machinery},
address = {New York, NY, USA},
url = {https://doi.org/10.1145/3491102.3501819},
doi = {10.1145/3491102.3501819},
booktitle = {Proceedings of the 2022 CHI Conference on Human Factors in Computing Systems},
articleno = {209},
numpages = {19},
location = {New Orleans, LA, USA},
series = {CHI '22}
}

@incollection{hart1988development,
  title     = {Development of {NASA}-{TLX} (Task Load Index): Results of Empirical and Theoretical Research},
  author    = {Hart, Sandra G. and Staveland, Lowell E.},
  booktitle = {Human Mental Workload},
  editor    = {Hancock, Peter A. and Meshkati, Najmedin},
  series    = {Advances in Psychology},
  volume    = {52},
  pages     = {139--183},
  year      = {1988},
  publisher = {North-Holland},
  address   = {Amsterdam, The Netherlands}
}

@article{listyani2019use,
  title={The Use of a Visual Image to Promote Narrative Writing Ability and Creativity.},
  author={Listyani, Lydia},
  journal={Eurasian Journal of Educational Research},
  volume={80},
  pages={193--223},
  year={2019},
  publisher={ERIC}
}

@article{mukramah2023effect,
  title={The Effect of Picture and Text Prompts on Idea Formulation and Organization of Descriptive Text.},
  author={Mukramah, Cut and Mustafa, Faisal and Sari, Diana Fauzia},
  journal={Indonesian Journal of English Language Teaching and Applied Linguistics},
  volume={7},
  number={2},
  pages={325--341},
  year={2023},
  publisher={ERIC}
}

@article{cherry2014quantifying,
  title={Quantifying the creativity support of digital tools through the creativity support index},
  author={Cherry, Erin and Latulipe, Celine},
  journal={ACM Transactions on Computer-Human Interaction (TOCHI)},
  volume={21},
  number={4},
  pages={1--25},
  year={2014},
  publisher={ACM New York, NY, USA}
}

@inproceedings{beaudouin2000instrumental,
author = {Beaudouin-Lafon, Michel},
title = {Instrumental interaction: an interaction model for designing post-WIMP user interfaces},
year = {2000},
isbn = {1581132166},
publisher = {Association for Computing Machinery},
address = {New York, NY, USA},
url = {https://doi.org/10.1145/332040.332473},
doi = {10.1145/332040.332473},
booktitle = {Proceedings of the SIGCHI Conference on Human Factors in Computing Systems},
pages = {446–453},
numpages = {8},
location = {The Hague, The Netherlands},
series = {CHI '00}
}

@inproceedings{10.1145/989863.989865,
author = {Beaudouin-Lafon, Michel},
title = {Designing interaction, not interfaces},
year = {2004},
isbn = {1581138679},
publisher = {Association for Computing Machinery},
address = {New York, NY, USA},
url = {https://doi-org.proxy.library.nd.edu/10.1145/989863.989865},
doi = {10.1145/989863.989865},
booktitle = {Proceedings of the Working Conference on Advanced Visual Interfaces},
pages = {15–22},
numpages = {8},
location = {Gallipoli, Italy},
series = {AVI '04}
}

@inproceedings{AIInstrumentsEmbodyingPrompts,
    author = {Riche, Nathalie and Offenwanger, Anna and Gmeiner, Frederic and Brown, David and Romat, Hugo and Pahud, Michel and Marquardt, Nicolai and Inkpen, Kori and Hinckley, Ken},
    title = {AI-Instruments: Embodying Prompts as Instruments to Abstract \& Reflect Graphical Interface Commands as General-Purpose Tools},
    year = {2025},
    isbn = {9798400713941},
    publisher = {Association for Computing Machinery},
    address = {New York, NY, USA},
    url = {https://doi.org/10.1145/3706598.3714259},
    doi = {10.1145/3706598.3714259},
    booktitle = {Proceedings of the 2025 CHI Conference on Human Factors in Computing Systems},
    articleno = {1104},
    numpages = {18},
    location = {    },
    series = {CHI '25}
}

@inproceedings{10.1145/3581641.3584095,
author = {Lawton, Tomas and Ibarrola, Francisco J and Ventura, Dan and Grace, Kazjon},
title = {Drawing with Reframer: Emergence and Control in Co-Creative\&nbsp;AI},
year = {2023},
isbn = {9798400701061},
publisher = {Association for Computing Machinery},
address = {New York, NY, USA},
url = {https://doi-org.proxy.library.nd.edu/10.1145/3581641.3584095},
doi = {10.1145/3581641.3584095},
booktitle = {Proceedings of the 28th International Conference on Intelligent User Interfaces},
pages = {264–277},
numpages = {14},
location = {Sydney, NSW, Australia},
series = {IUI '23}
}

@misc{darejeh2024criticalanalysiscognitiveload,
      title={A critical analysis of cognitive load measurement methods for evaluating the usability of different types of interfaces: guidelines and framework for Human-Computer Interaction}, 
      author={Ali Darejeh and Nadine Marcusa and Gelareh Mohammadi and John Sweller},
      year={2024},
      eprint={2402.11820},
      archivePrefix={arXiv},
      primaryClass={cs.HC},
      url={https://arxiv.org/abs/2402.11820}, 
}

@inproceedings{10.1145/3635636.3656201,
author = {Chakrabarty, Tuhin and Padmakumar, Vishakh and Brahman, Faeze and Muresan, Smaranda},
title = {Creativity Support in the Age of Large Language Models: An Empirical Study Involving Professional Writers},
year = {2024},
isbn = {9798400704857},
publisher = {Association for Computing Machinery},
address = {New York, NY, USA},
url = {https://doi-org.proxy.library.nd.edu/10.1145/3635636.3656201},
doi = {10.1145/3635636.3656201},
booktitle = {Proceedings of the 16th Conference on Creativity \& Cognition},
pages = {132–155},
numpages = {24},
location = {Chicago, IL, USA},
series = {C\&C '24}
}

@article{doyle1998writer,
  title={The writer tells: The creative process in the writing of literary fiction},
  author={Doyle, Charlotte L},
  journal={Creativity Research Journal},
  volume={11},
  number={1},
  pages={29--37},
  year={1998},
  publisher={Taylor \& Francis}
}

@article{bal2013does,
  title={How does fiction reading influence empathy? An experimental investigation on the role of emotional transportation},
  author={Bal, P Matthijs and Veltkamp, Martijn},
  journal={PloS one},
  volume={8},
  number={1},
  pages={e55341},
  year={2013},
  publisher={Public Library of Science San Francisco, USA}
}

@inproceedings{geroDesignSpaceWriting2022,
  title = {A {{Design Space}} for {{Writing Support Tools Using}} a {{Cognitive Process Model}} of {{Writing}}},
  booktitle = {Proceedings of the {{First Workshop}} on {{Intelligent}} and {{Interactive Writing Assistants}} ({{In2Writing}} 2022)},
  author = {Gero, Katy and Calderwood, Alex and Li, Charlotte and Chilton, Lydia},
  year = {2022},
  pages = {11--24},
  publisher = {Association for Computational Linguistics},
  address = {Dublin, Ireland},
  doi = {10.18653/v1/2022.in2writing-1.2},
  urldate = {2024-10-22},
  langid = {english}
}

@inproceedings{geroSocialDynamicsAI2023,
    author = {Gero, Katy Ilonka and Long, Tao and Chilton, Lydia B},
    title = {Social Dynamics of AI Support in Creative Writing},
    year = {2023},
    isbn = {9781450394215},
    publisher = {Association for Computing Machinery},
    address = {New York, NY, USA},
    url = {https://doi.org/10.1145/3544548.3580782},
    doi = {10.1145/3544548.3580782},
    booktitle = {Proceedings of the 2023 CHI Conference on Human Factors in Computing Systems},
    articleno = {245},
    numpages = {15},
    location = {Hamburg, Germany},
    series = {CHI '23}
}

@inproceedings{shi2025brickify,
author = {Shi, Xinyu and Wang, Yinghou and Rossi, Ryan and Zhao, Jian},
title = {Brickify: Enabling Expressive Design Intent Specification through Direct Manipulation on Design Tokens},
year = {2025},
isbn = {9798400713941},
publisher = {Association for Computing Machinery},
address = {New York, NY, USA},
url = {https://doi.org/10.1145/3706598.3714087},
doi = {10.1145/3706598.3714087},
booktitle = {Proceedings of the 2025 CHI Conference on Human Factors in Computing Systems},
articleno = {424},
numpages = {20},
location = {},
series = {CHI '25}
}

@book{burroway2022writing,
  author    = {Janet Burroway and Elizabeth Stuckey-French and Ned Stuckey-French},
  title     = {Writing Fiction: A Guide to Narrative Craft},
  edition   = {10th},
  year      = {2019},
  publisher = {University of Chicago Press},
  address   = {Chicago, IL, USA},
}

@book{keen2007empathy,
  author    = {Suzanne Keen},
  title     = {Empathy and the Novel},
  year      = {2007},
  publisher = {Oxford University Press},
  address   = {Oxford, UK},
  doi       = {10.1093/acprof:oso/9780195175769.001.0001},
  isbn      = {9780195175769}
}

@book{genette1980narrative,
  author    = {G{\'e}rard Genette},
  title     = {Narrative Discourse: An Essay in Method},
  translator= {Jane E. Lewin},
  year      = {1980},
  publisher = {Cornell University Press},
  address   = {Ithaca, NY},
  pages     = {285},
  isbn      = {9780801492594}
}

@article{flower1981cognitive,
  title={A cognitive process theory of writing},
  author={Flower, Linda and Hayes, John R},
  journal={College Composition \& Communication},
  volume={32},
  number={4},
  pages={365--387},
  year={1981},
  publisher={NCTE}
}

@inproceedings{xia2018spacetime,
author = {Xia, Haijun and Herscher, Sebastian and Perlin, Ken and Wigdor, Daniel},
title = {Spacetime: Enabling Fluid Individual and Collaborative Editing in Virtual Reality},
year = {2018},
isbn = {9781450359481},
publisher = {Association for Computing Machinery},
address = {New York, NY, USA},
url = {https://doi.org/10.1145/3242587.3242597},
doi = {10.1145/3242587.3242597},
booktitle = {Proceedings of the 31st Annual ACM Symposium on User Interface Software and Technology},
pages = {853–866},
numpages = {14},
location = {Berlin, Germany},
series = {UIST '18}
}

@phdthesis{masson2023transforming,
  title={Transforming the Reading Experience of Scientific Documents with Polymorphism},
  author={Masson, Damien},
  year={2023},
  school={University of Waterloo}
}

@article{angelini2015move,
  title={Move, hold and touch: a framework for tangible gesture interactive systems},
  author={Angelini, Leonardo and Lalanne, Denis and Van den Hoven, Elise and Abou Khaled, Omar and Mugellini, Elena},
  journal={Machines},
  volume={3},
  number={3},
  pages={173--207},
  year={2015},
  publisher={MDPI}
}

@article{kent2000conceptual,
  title={Conceptual knowledge markup language: An introduction},
  author={Kent, Robert E},
  journal={Netnomics},
  volume={2},
  number={2},
  pages={139--169},
  year={2000},
  publisher={Springer}
}

@inproceedings{yen2023coladder,
author = {Yen, Ryan and Zhu, Jiawen Stefanie and Suh, Sangho and Xia, Haijun and Zhao, Jian},
title = {CoLadder: Manipulating Code Generation via Multi-Level Blocks},
year = {2024},
isbn = {9798400706288},
publisher = {Association for Computing Machinery},
address = {New York, NY, USA},
url = {https://doi.org/10.1145/3654777.3676357},
doi = {10.1145/3654777.3676357},
booktitle = {Proceedings of the 37th Annual ACM Symposium on User Interface Software and Technology},
articleno = {11},
numpages = {20},
location = {Pittsburgh, PA, USA},
series = {UIST '24}
}

@inproceedings{suh2024luminate,
author = {Suh, Sangho and Chen, Meng and Min, Bryan and Li, Toby Jia-Jun and Xia, Haijun},
title = {Luminate: Structured Generation and Exploration of Design Space with Large Language Models for Human-AI Co-Creation},
year = {2024},
isbn = {9798400703300},
publisher = {Association for Computing Machinery},
address = {New York, NY, USA},
url = {https://doi-org.proxy.library.nd.edu/10.1145/3613904.3642400},
doi = {10.1145/3613904.3642400},
booktitle = {Proceedings of the 2024 CHI Conference on Human Factors in Computing Systems},
articleno = {644},
numpages = {26},
location = {Honolulu, HI, USA},
series = {CHI '24}
}

@article{miall1999literariness,
  title={What is literariness? Three components of literary reading},
  author={Miall, David S and Kuiken, Don},
  journal={Discourse processes},
  volume={28},
  number={2},
  pages={121--138},
  year={1999},
  publisher={Taylor \& Francis}
}

@article{arbib1992schema,
  title={Schema theory},
  author={Arbib, Michael A},
  journal={The encyclopedia of artificial intelligence},
  volume={2},
  pages={1427--1443},
  year={1992},
  publisher={Wiley Interscience New York, NY}
}

@article{green2000role,
  title={The role of transportation in the persuasiveness of public narratives.},
  author={Green, Melanie C and Brock, Timothy C},
  journal={Journal of personality and social psychology},
  volume={79},
  number={5},
  pages={701},
  year={2000},
  publisher={American Psychological Association}
}

@inproceedings{10.1145/3706598.3713862,
author = {Masson, Damien and Kim, Young-Ho and Chevalier, Fanny},
title = {Textoshop: Interactions Inspired by Drawing Software to Facilitate Text Editing},
year = {2025},
isbn = {9798400713941},
publisher = {Association for Computing Machinery},
address = {New York, NY, USA},
url = {https://doi.org/10.1145/3706598.3713862},
doi = {10.1145/3706598.3713862},
booktitle = {Proceedings of the 2025 CHI Conference on Human Factors in Computing Systems},
articleno = {1087},
numpages = {14},
location = {
},
series = {CHI '25}
}

@article{thornberg2012informed,
  title={Informed grounded theory},
  author={Thornberg, Robert},
  journal={Scandinavian journal of educational research},
  volume={56},
  number={3},
  pages={243--259},
  year={2012},
  publisher={Taylor \& Francis}
}

@inproceedings{chung2024patchview,
author = {Chung, John Joon Young and Kreminski, Max},
title = {Patchview: LLM-powered Worldbuilding with Generative Dust and Magnet Visualization},
year = {2024},
isbn = {9798400706288},
publisher = {Association for Computing Machinery},
address = {New York, NY, USA},
url = {https://doi.org/10.1145/3654777.3676352},
doi = {10.1145/3654777.3676352},
booktitle = {Proceedings of the 37th Annual ACM Symposium on User Interface Software and Technology},
articleno = {77},
numpages = {19},
location = {Pittsburgh, PA, USA},
series = {UIST '24}
}

@inproceedings{rosenbergDrawTalkingBuildingInteractive2024,
author = {Rosenberg, Karl Toby and Kazi, Rubaiat Habib and Wei, Li-Yi and Xia, Haijun and Perlin, Ken},
title = {DrawTalking: Building Interactive Worlds by Sketching and Speaking},
year = {2024},
isbn = {9798400706288},
publisher = {Association for Computing Machinery},
address = {New York, NY, USA},
url = {https://doi.org/10.1145/3654777.3676334},
doi = {10.1145/3654777.3676334},
booktitle = {Proceedings of the 37th Annual ACM Symposium on User Interface Software and Technology},
articleno = {76},
numpages = {25},
location = {Pittsburgh, PA, USA},
series = {UIST '24}
}

@inproceedings{wang2025aideation,
author = {Wang, Wen-Fan and Lu, Chien-Ting and Ponsa i Campany\`{a}, Nil and Chen, Bing-Yu and Chen, Mike Y.},
title = {AIdeation: Designing a Human-AI Collaborative Ideation System for Concept Designers},
year = {2025},
isbn = {9798400713941},
publisher = {Association for Computing Machinery},
address = {New York, NY, USA},
url = {https://doi.org/10.1145/3706598.3714148},
doi = {10.1145/3706598.3714148},
booktitle = {Proceedings of the 2025 CHI Conference on Human Factors in Computing Systems},
articleno = {21},
numpages = {28},
location = {
},
series = {CHI '25}
}

@article{kanellopoulou2019dual,
  title={The dual-coding and multimedia learning theories: Film subtitles as a vocabulary teaching tool},
  author={Kanellopoulou, Catherine and Kermanidis, Katia Lida and Giannakoulopoulos, Andreas},
  journal={Education Sciences},
  volume={9},
  number={3},
  pages={210},
  year={2019},
  publisher={MDPI}
}

@article{rezaee2011investigating,
  author  = {Abbas Ali Rezaee and Neda Sharbaf Shoar},
  title   = {Investigating the Effect of Using Multiple Sensory Modes of Glossing Vocabulary Items in a Reading Text with Multimedia Annotations},
  journal = {English Language Teaching},
  volume  = {4},
  number  = {2},
  pages   = {25--34},
  year    = {2011},
  doi     = {10.5539/elt.v4n2p25}
}

@article{huang2024sight,
  title={From sight to insight: A multi-task approach with the visual language decoding model},
  author={Huang, Wei and Yang, Pengfei and Tang, Ying and Qin, Fan and Li, Hengjiang and Wu, Diwei and Ren, Wei and Wang, Sizhuo and Li, Jingpeng and Zhu, Yucheng and others},
  journal={Information Fusion},
  volume={112},
  pages={102573},
  year={2024},
  publisher={Elsevier}
}

@inproceedings{10.1145/3746058.3758469,
author = {Long, Tao and Wang, Sitong and Fabre, \'{E}milie and Wang, Tony and Sathya, Anup and Wu, Jason and Petridis, Savvas Dimitrios and Li, Ding and Chakrabarty, Tuhin and Jiang, Yue and Li, Jingyi and Tseng, Tiffany and Nakagaki, Ken and Yang, Qian and Martelaro, Nikolas and Nickerson, Jeffrey V and Chilton, Lydia B},
title = {Facilitating Longitudinal Interaction Studies of AI Systems},
year = {2025},
isbn = {9798400720369},
publisher = {Association for Computing Machinery},
address = {New York, NY, USA},
url = {https://doi-org.proxy.library.nd.edu/10.1145/3746058.3758469},
doi = {10.1145/3746058.3758469},
booktitle = {Adjunct Proceedings of the 38th Annual ACM Symposium on User Interface Software and Technology},
articleno = {13},
numpages = {5},
location = {
},
series = {UIST Adjunct '25}
}

@inproceedings{10.1145/302979.303030,
author = {Horvitz, Eric},
title = {Principles of mixed-initiative user interfaces},
year = {1999},
isbn = {0201485591},
publisher = {Association for Computing Machinery},
address = {New York, NY, USA},
url = {https://doi-org.proxy.library.nd.edu/10.1145/302979.303030},
doi = {10.1145/302979.303030},
booktitle = {Proceedings of the SIGCHI Conference on Human Factors in Computing Systems},
pages = {159–166},
numpages = {8},
location = {Pittsburgh, Pennsylvania, USA},
series = {CHI '99}
}

@misc{kwan2024mt,
      title={MT-Eval: A Multi-Turn Capabilities Evaluation Benchmark for Large Language Models}, 
      author={Wai-Chung Kwan and Xingshan Zeng and Yuxin Jiang and Yufei Wang and Liangyou Li and Lifeng Shang and Xin Jiang and Qun Liu and Kam-Fai Wong},
      year={2024},
      eprint={2401.16745},
      archivePrefix={arXiv},
      primaryClass={cs.CL},
      url={https://arxiv.org/abs/2401.16745}, 
}

@inproceedings{fu2024being,
author = {Fu, Kexue and Wu, Ruishan and Tang, Yuying and Chen, Yixin and Liu, Bowen and LC, RAY},
title = {"Being Eroded, Piece by Piece": Enhancing Engagement and Storytelling in Cultural Heritage Dissemination by Exhibiting GenAI Co-Creation Artifacts},
year = {2024},
isbn = {9798400705830},
publisher = {Association for Computing Machinery},
address = {New York, NY, USA},
url = {https://doi.org/10.1145/3643834.3660711},
doi = {10.1145/3643834.3660711},
booktitle = {Proceedings of the 2024 ACM Designing Interactive Systems Conference},
pages = {2833–2850},
numpages = {18},
location = {Copenhagen, Denmark},
series = {DIS '24}
}

@inproceedings{10.1145/3715336.3735766,
author = {Bourgault, Samuelle and Wei, Li-Yi and Jacobs, Jennifer and Kazi, Rubaiat Habib},
title = {Narrative Motion Blocks: Combining Direct Manipulation and Natural Language Interactions for Animation Creation},
year = {2025},
isbn = {9798400714856},
publisher = {Association for Computing Machinery},
address = {New York, NY, USA},
url = {https://doi-org.proxy.library.nd.edu/10.1145/3715336.3735766},
doi = {10.1145/3715336.3735766},
booktitle = {Proceedings of the 2025 ACM Designing Interactive Systems Conference},
pages = {1366–1386},
numpages = {21},
location = {
},
series = {DIS '25}
}

@article{clark1991dual,
  title={Dual coding theory and education},
  author={Clark, James M and Paivio, Allan},
  journal={Educational psychology review},
  volume={3},
  number={3},
  pages={149--210},
  year={1991},
  publisher={Springer}
}

@online{ChatGPT,
  title = {ChatGPT},
  url = {https://chatgpt.com/c/552d76a2-93a8-45d0-a883-232346937ce7},
  urldate = {2024-07-30},
  langid = {auto}
}

@inproceedings{10.1145/3746059.3747772,
author = {Suh, Sangho and Lai, Michael and Pu, Kevin and Dow, Steven P. and Grossman, Tovi},
title = {StoryEnsemble: Enabling Dynamic Exploration \& Iteration in the Design Process with AI and Forward-Backward Propagation},
year = {2025},
isbn = {9798400720376},
publisher = {Association for Computing Machinery},
address = {New York, NY, USA},
url = {https://doi-org.proxy.library.nd.edu/10.1145/3746059.3747772},
doi = {10.1145/3746059.3747772},
booktitle = {Proceedings of the 38th Annual ACM Symposium on User Interface Software and Technology},
articleno = {203},
numpages = {36},
location = {
},
series = {UIST '25}
}

@inproceedings{choi2024creativeconnect,
author = {Choi, DaEun and Hong, Sumin and Park, Jeongeon and Chung, John Joon Young and Kim, Juho},
title = {CreativeConnect: Supporting Reference Recombination for Graphic Design Ideation with Generative AI},
year = {2024},
isbn = {9798400703300},
publisher = {Association for Computing Machinery},
address = {New York, NY, USA},
url = {https://doi.org/10.1145/3613904.3642794},
doi = {10.1145/3613904.3642794},
booktitle = {Proceedings of the 2024 CHI Conference on Human Factors in Computing Systems},
articleno = {1055},
numpages = {25},
location = {Honolulu, HI, USA},
series = {CHI '24}
}

@inproceedings{chungToytellerAIpoweredVisual2025a,
author = {Chung, John Joon Young and Roemmele, Melissa and Kreminski, Max},
title = {Toyteller: AI-powered Visual Storytelling Through Toy-Playing with Character Symbols},
year = {2025},
isbn = {9798400713941},
publisher = {Association for Computing Machinery},
address = {New York, NY, USA},
url = {https://doi.org/10.1145/3706598.3713435},
doi = {10.1145/3706598.3713435},
booktitle = {Proceedings of the 2025 CHI Conference on Human Factors in Computing Systems},
articleno = {331},
numpages = {23},
location = {
},
series = {CHI '25}
}

@inproceedings{ClueCartSupportingGame,
author = {Wang, Xiyuan and Cao, Yi-Fan and Xiong, Junjie and Chen, Sizhe and Li, Wenxuan and Zhang, Junjie and Li, Quan},
title = {ClueCart: Supporting Game Story Interpretation and Narrative Inference from Fragmented Clues},
year = {2025},
isbn = {9798400713941},
publisher = {Association for Computing Machinery},
address = {New York, NY, USA},
url = {https://doi-org.proxy.library.nd.edu/10.1145/3706598.3713381},
doi = {10.1145/3706598.3713381},
booktitle = {Proceedings of the 2025 CHI Conference on Human Factors in Computing Systems},
articleno = {410},
numpages = {26},
location = {
},
series = {CHI '25}
}

@article{barsalou2008grounded,
  title={Grounded cognition},
  author={Barsalou, Lawrence W},
  journal={Annu. Rev. Psychol.},
  volume={59},
  number={1},
  pages={617--645},
  year={2008},
  publisher={Annual Reviews}
}

@inproceedings{dangWorldSmithIterativeExpressive2023,
author = {Dang, Hai and Brudy, Frederik and Fitzmaurice, George and Anderson, Fraser},
title = {WorldSmith: Iterative and Expressive Prompting for World Building with a Generative AI},
year = {2023},
isbn = {9798400701320},
publisher = {Association for Computing Machinery},
address = {New York, NY, USA},
url = {https://doi.org/10.1145/3586183.3606772},
doi = {10.1145/3586183.3606772},
booktitle = {Proceedings of the 36th Annual ACM Symposium on User Interface Software and Technology},
articleno = {63},
numpages = {17},
location = {San Francisco, CA, USA},
series = {UIST '23}
}

@inproceedings{lin2025inkspire,
author = {Lin, David Chuan-En and Kang, Hyeonsu B. and Martelaro, Nikolas and Kittur, Aniket and Chen, Yan-Ying and Hong, Matthew K.},
title = {Inkspire: Supporting Design Exploration with Generative AI through Analogical Sketching},
year = {2025},
isbn = {9798400713941},
publisher = {Association for Computing Machinery},
address = {New York, NY, USA},
url = {https://doi.org/10.1145/3706598.3713397},
doi = {10.1145/3706598.3713397},
booktitle = {Proceedings of the 2025 CHI Conference on Human Factors in Computing Systems},
articleno = {427},
numpages = {18},
location = {
},
series = {CHI '25}
}

@misc{park2024character,
      title={A Character-Centric Creative Story Generation via Imagination}, 
      author={Kyeongman Park and Minbeom Kim and Kyomin Jung},
      year={2024},
      eprint={2409.16667},
      archivePrefix={arXiv},
      primaryClass={cs.CL},
      url={https://arxiv.org/abs/2409.16667}, 
}

@inproceedings{10.1145/3746059.3747758,
author = {Masson, Damien and Zhao, Zixin and Chevalier, Fanny},
title = {Visual Story-Writing: Writing by Manipulating Visual Representations of Stories},
year = {2025},
isbn = {9798400720376},
publisher = {Association for Computing Machinery},
address = {New York, NY, USA},
url = {https://doi.org/10.1145/3746059.3747758},
doi = {10.1145/3746059.3747758},
booktitle = {Proceedings of the 38th Annual ACM Symposium on User Interface Software and Technology},
articleno = {70},
numpages = {15},
location = {
},
series = {UIST '25}
}

@article{brown2021neural,
  title={The neural basis of creative production: A cross-modal ALE meta-analysis},
  author={Brown, Steven and Kim, Eunseon},
  journal={Open Psychology},
  volume={3},
  number={1},
  pages={103--132},
  year={2021},
  publisher={Sciendo}
}

@inproceedings{mishraWhatIFBranchedNarrative2025,
author = {Mishra, Aditi and Brudy, Frederik and Zhou, Qian and Fitzmaurice, George and Anderson, Fraser},
title = {WhatIF: Branched Narrative Fiction Visualization for Authoring Emergent Narratives using Large Language Models},
year = {2025},
isbn = {9798400712890},
publisher = {Association for Computing Machinery},
address = {New York, NY, USA},
url = {https://doi.org/10.1145/3698061.3726933},
doi = {10.1145/3698061.3726933},
booktitle = {Proceedings of the 2025 Conference on Creativity and Cognition},
pages = {590–605},
numpages = {16},
location = {
},
series = {C\&C '25}
}

@inproceedings{raoScriptVizVisualizationTool2024,
author = {Rao, Anyi and Chou, Jean-Pe\"{\i}c and Agrawala, Maneesh},
title = {ScriptViz: A Visualization Tool to Aid Scriptwriting based on a Large Movie Database},
year = {2024},
isbn = {9798400706288},
publisher = {Association for Computing Machinery},
address = {New York, NY, USA},
url = {https://doi.org/10.1145/3654777.3676402},
doi = {10.1145/3654777.3676402},
booktitle = {Proceedings of the 37th Annual ACM Symposium on User Interface Software and Technology},
articleno = {21},
numpages = {13},
location = {Pittsburgh, PA, USA},
series = {UIST '24}
}

@misc{MiroWebsite,
  author       = {{Miro}},
  title        = {Miro – AI Innovation Workspace},
  year         = {2025},
  howpublished = {\url{https://miro.com/}},
  note         = {Accessed: 2025-12-01}
}

@inproceedings{10.1145/3613904.3642731,
author = {Chakrabarty, Tuhin and Laban, Philippe and Agarwal, Divyansh and Muresan, Smaranda and Wu, Chien-Sheng},
title = {Art or Artifice? Large Language Models and the False Promise of Creativity},
year = {2024},
isbn = {9798400703300},
publisher = {Association for Computing Machinery},
address = {New York, NY, USA},
url = {https://doi-org.proxy.library.nd.edu/10.1145/3613904.3642731},
doi = {10.1145/3613904.3642731},
booktitle = {Proceedings of the 2024 CHI Conference on Human Factors in Computing Systems},
articleno = {30},
numpages = {34},
location = {Honolulu, HI, USA},
series = {CHI '24}
}

@article{doshi2024generative,
  title={Generative AI enhances individual creativity but reduces the collective diversity of novel content},
  author={Doshi, Anil R and Hauser, Oliver P},
  journal={Science advances},
  volume={10},
  number={28},
  pages={eadn5290},
  year={2024},
  publisher={American Association for the Advancement of Science}
}

@inproceedings{10.1145/3613904.3642625,
author = {Li, Zhuoyan and Liang, Chen and Peng, Jing and Yin, Ming},
title = {The Value, Benefits, and Concerns of Generative AI-Powered Assistance in Writing},
year = {2024},
isbn = {9798400703300},
publisher = {Association for Computing Machinery},
address = {New York, NY, USA},
url = {https://doi-org.proxy.library.nd.edu/10.1145/3613904.3642625},
doi = {10.1145/3613904.3642625},
booktitle = {Proceedings of the 2024 CHI Conference on Human Factors in Computing Systems},
articleno = {1048},
numpages = {25},
location = {Honolulu, HI, USA},
series = {CHI '24}
}

@article{kirsh2010thinking,
  title={Thinking with external representations},
  author={Kirsh, David},
  journal={AI \& society},
  volume={25},
  number={4},
  pages={441--454},
  year={2010},
  publisher={Springer}
}

@article{10.1145/3511599,
author = {Singh, Nikhil and Bernal, Guillermo and Savchenko, Daria and Glassman, Elena L.},
title = {Where to Hide a Stolen Elephant: Leaps in Creative Writing with Multimodal Machine Intelligence},
year = {2023},
issue_date = {October 2023},
publisher = {Association for Computing Machinery},
address = {New York, NY, USA},
volume = {30},
number = {5},
issn = {1073-0516},
url = {https://doi-org.proxy.library.nd.edu/10.1145/3511599},
doi = {10.1145/3511599},
journal = {ACM Trans. Comput.-Hum. Interact.},
month = sep,
articleno = {68},
numpages = {57}
}

@inproceedings{wangPlotMapAutomatedLayout2024,
  author       = {Wang, Yi and Luo, Jieliang and Gaier, Adam and Atherton, Evan and Koch, Hilmar},
  title        = {PlotMap: Automated Layout Design for Building Game Worlds},
  booktitle    = {Proceedings of the 2024 IEEE Conference on Games (CoG)},
  series       = {CoG '24},
  year         = {2024},
  pages        = {1--8},
  publisher    = {IEEE},
  organization = {Institute of Electrical and Electronics Engineers},
  address      = {Milan, Italy},
  doi          = {10.1109/CoG60054.2024.10645627},
  isbn         = {979-8-3503-5067-8}
}

@article{Sidra19082025,
author = {Sidra Sidra and Claire Mason},
title = {Generative AI in Human-AI Collaboration: Validation of the Collaborative AI Literacy and Collaborative AI Metacognition Scales for Effective Use},
journal = {International Journal of Human–Computer Interaction},
volume = {0},
number = {0},
pages = {1--25},
year = {2025},
publisher = {Taylor \& Francis},
doi = {10.1080/10447318.2025.2543997},
URL = {    
        https://doi.org/10.1080/10447318.2025.2543997
=
},
eprint = {    
        https://doi.org/10.1080/10447318.2025.2543997
}

}

@article{Rafner01122025,
author = {Janet Rafner and Blanka Zana and Ida Bang Hansen and Simon Ceh and Jacob Sherson and Mathias Benedek and Izabela Lebuda},
title = {Agency in Human-AI Collaboration for Image Generation and Creative Writing: Preliminary Insights from Think-Aloud Protocols},
journal = {Creativity Research Journal},
volume = {0},
number = {0},
pages = {1--24},
year = {2025},
publisher = {Routledge},
doi = {10.1080/10400419.2025.2587803},


URL = { 
    
        https://doi.org/10.1080/10400419.2025.2587803
    
    

},
eprint = { 
    
        https://doi.org/10.1080/10400419.2025.2587803
    
    

}

}

@inproceedings{10.1145/3639701.3656325,
author = {Weber, Christoph Johannes and Burgkart, Sebastian and Rothe, Sylvia},
title = {wr-AI-ter: Enhancing Ownership Perception in AI-Driven Script Writing},
year = {2024},
isbn = {9798400705038},
publisher = {Association for Computing Machinery},
address = {New York, NY, USA},
url = {https://doi.org/10.1145/3639701.3656325},
doi = {10.1145/3639701.3656325},
booktitle = {Proceedings of the 2024 ACM International Conference on Interactive Media Experiences},
pages = {145–156},
numpages = {12},
location = {Stockholm, Sweden},
series = {IMX '24}
}

@article{10.1145/3637875,
author = {Draxler, Fiona and Werner, Anna and Lehmann, Florian and Hoppe, Matthias and Schmidt, Albrecht and Buschek, Daniel and Welsch, Robin},
title = {The AI Ghostwriter Effect: When Users do not Perceive Ownership of AI-Generated Text but Self-Declare as Authors},
year = {2024},
issue_date = {April 2024},
publisher = {Association for Computing Machinery},
address = {New York, NY, USA},
volume = {31},
number = {2},
issn = {1073-0516},
url = {https://doi.org/10.1145/3637875},
doi = {10.1145/3637875},
journal = {ACM Trans. Comput.-Hum. Interact.},
month = feb,
articleno = {25},
numpages = {40}
}

@misc{wang2025script2screen,
      title={Script2Screen: Supporting Dialogue Scriptwriting with Interactive Audiovisual Generation}, 
      author={Zhecheng Wang and Jiaju Ma and Eitan Grinspun and Tovi Grossman and Bryan Wang},
      year={2025},
      eprint={2504.14776},
      archivePrefix={arXiv},
      primaryClass={cs.HC},
      url={https://arxiv.org/abs/2504.14776}, 
}

@inproceedings{yanXCreationGraphbasedCrossmodal2023,
author = {Yan, Zihan and Yang, Chunxu and Liang, Qihao and Chen, Xiang 'Anthony'},
title = {XCreation: A Graph-based Crossmodal Generative Creativity Support Tool},
year = {2023},
isbn = {9798400701320},
publisher = {Association for Computing Machinery},
address = {New York, NY, USA},
url = {https://doi.org/10.1145/3586183.3606826},
doi = {10.1145/3586183.3606826},
booktitle = {Proceedings of the 36th Annual ACM Symposium on User Interface Software and Technology},
articleno = {48},
numpages = {15},
location = {San Francisco, CA, USA},
series = {UIST '23}
}

@inproceedings{yenCodeShapingIterative2024,
author = {Yen, Ryan and Zhao, Jian and Vogel, Daniel},
title = {Code Shaping: Iterative Code Editing with Free-form AI-Interpreted Sketching},
year = {2025},
isbn = {9798400713941},
publisher = {Association for Computing Machinery},
address = {New York, NY, USA},
url = {https://doi.org/10.1145/3706598.3713822},
doi = {10.1145/3706598.3713822},
booktitle = {Proceedings of the 2025 CHI Conference on Human Factors in Computing Systems},
articleno = {872},
numpages = {17},
location = {
},
series = {CHI '25}
}

@book{booker2004seven,
  author    = {Christopher Booker},
  title     = {The Seven Basic Plots: Why We Tell Stories},
  year      = {2004},
  publisher = {Continuum},
  address   = {London; },
  pages     = {728},
  isbn      = {9780826452092}
}

@misc{claude_new,
  author       = {Anthropic},
  title        = {Claude 3.5 Sonnet},
  year         = {2024},
  howpublished = {\url{https://www.anthropic.com/news/claude-3-5-sonnet}},
  note         = {Claude 3.5 Sonnet large language model used via \url{https://claude.ai}, June 2025.}
}

@online{midjourney_home,
  title        = {Midjourney: Text-to-Image Generation System (Version 6.1)},
  organization = {Midjourney, Inc.},
  year         = {2024},
  url          = {https://www.midjourney.com/},
  urldate      = {2025-06-16},
  note         = {Midjourney v6.1, the default text-to-image model until June 16, 2025, accessed via the Midjourney web/Discord service.}
}

@online{openai_hello_gpt4o,
  organization = {OpenAI},
  title        = {GPT-4o},
  year         = {2024},
  howpublished = {\url{https://openai.com/index/hello-gpt-4o/}},
  note         = {Large multimodal model (GPT-4o; ChatGPT model ID \texttt{chatgpt-4o-latest}) used via \url{https://chatgpt.com}, June 2025.}
}

@online{figma_home,
  title        = {Figma: Collaborative Interface Design Tool},
  organization = {Figma, Inc.},
  year         = {2025},
  url          = {https://www.figma.com/},
  urldate      = {2025-08-29},
  note         = {Official homepage describing Figma’s collaborative design, prototyping, whiteboarding, presentation, and AI features}
}

@misc{openai_gpt4o_api,
  author       = {OpenAI},
  title        = {GPT-4o},
  year         = {2024},
  howpublished = {\url{https://platform.openai.com/docs/models/gpt-4o}},
  note         = {Flagship multimodal model ``GPT-4o''. Used via the OpenAI API with model ID \texttt{gpt-4o} (and deployment aliases such as \texttt{chatgpt-4o-latest}) in August 2025.}
}

@misc{openai_o4_mini_api,
  author       = {OpenAI},
  title        = {o4-mini},
  year         = {2025},
  howpublished = {\url{https://platform.openai.com/docs/models}},
  note         = {OpenAI ``o4-mini'' model from the o-series. Used via the OpenAI API with model ID \texttt{o4-mini} in August 2025.}
}

@misc{bfl_flux1_api,
  author       = {{Black Forest Labs}},
  title        = {FLUX.1: Next-generation text-to-image diffusion models},
  year         = {2024},
  howpublished = {\url{https://bfl.ai/}},
  note         = {FLUX.1 diffusion model family (e.g., FLUX.1 [pro] / [dev] / [schnell]) accessed via the Black Forest Labs API in August 2025.}
}

@inproceedings{lee2024design,
author = {Lee, Mina and Gero, Katy Ilonka and Chung, John Joon Young and Shum, Simon Buckingham and Raheja, Vipul and Shen, Hua and Venugopalan, Subhashini and Wambsganss, Thiemo and Zhou, David and Alghamdi, Emad A. and August, Tal and Bhat, Avinash and Choksi, Madiha Zahrah and Dutta, Senjuti and Guo, Jin L.C. and Hoque, Md Naimul and Kim, Yewon and Knight, Simon and Neshaei, Seyed Parsa and Shibani, Antonette and Shrivastava, Disha and Shroff, Lila and Sergeyuk, Agnia and Stark, Jessi and Sterman, Sarah and Wang, Sitong and Bosselut, Antoine and Buschek, Daniel and Chang, Joseph Chee and Chen, Sherol and Kreminski, Max and Park, Joonsuk and Pea, Roy and Rho, Eugenia Ha Rim and Shen, Zejiang and Siangliulue, Pao},
title = {A Design Space for Intelligent and Interactive Writing Assistants},
year = {2024},
isbn = {9798400703300},
publisher = {Association for Computing Machinery},
address = {New York, NY, USA},
url = {https://doi.org/10.1145/3613904.3642697},
doi = {10.1145/3613904.3642697},
booktitle = {Proceedings of the 2024 CHI Conference on Human Factors in Computing Systems},
articleno = {1054},
numpages = {35},
location = {Honolulu, HI, USA},
series = {CHI '24}
}

@inproceedings{zhaoMakingWriteConnections2025,
author = {Zhao, Zixin and Masson, Damien and Kim, Young-Ho and Penn, Gerald and Chevalier, Fanny},
title = {Making the Write Connections: Linking Writing Support Tools with Writer Needs},
year = {2025},
isbn = {9798400713941},
publisher = {Association for Computing Machinery},
address = {New York, NY, USA},
url = {https://doi.org/10.1145/3706598.3713161},
doi = {10.1145/3706598.3713161},
booktitle = {Proceedings of the 2025 CHI Conference on Human Factors in Computing Systems},
articleno = {1216},
numpages = {21},
location = {
},
series = {CHI '25}
}

@article{legaspi_synthetic_2019,
	title = {Synthetic agency: sense of agency in artificial intelligence},
	volume = {29},
	issn = {23521546},
	shorttitle = {Synthetic agency},
	url = {https://linkinghub.elsevier.com/retrieve/pii/S2352154618301700},
	doi = {10.1016/j.cobeha.2019.04.004},
	language = {en},
	urldate = {2025-11-22},
	journal = {Current Opinion in Behavioral Sciences},
	author = {Legaspi, Roberto and He, Zhengqi and Toyoizumi, Taro},
	month = oct,
	year = {2019},
	pages = {84--90},
}

@article{10.1145/3677102,
author = {Ling, Long and Chen, Xinyi and Wen, Ruoyu and Li, Toby Jia-Jun and LC, RAY},
title = {Sketchar: Supporting Character Design and Illustration Prototyping Using Generative AI},
year = {2024},
issue_date = {October 2024},
publisher = {Association for Computing Machinery},
address = {New York, NY, USA},
volume = {8},
number = {CHI PLAY},
url = {https://doi.org/10.1145/3677102},
doi = {10.1145/3677102},
journal = {Proc. ACM Hum.-Comput. Interact.},
month = oct,
articleno = {337},
numpages = {28}
}

@article{draxler_ai_2024,
	title = {The {AI} {Ghostwriter} {Effect}: {When} {Users} do not {Perceive} {Ownership} of {AI}-{Generated} {Text} but {Self}-{Declare} as {Authors}},
	volume = {31},
	issn = {1073-0516, 1557-7325},
	shorttitle = {The {AI} {Ghostwriter} {Effect}},
	url = {https://dl.acm.org/doi/10.1145/3637875},
	doi = {10.1145/3637875},
	language = {en},
	number = {2},
	urldate = {2025-11-22},
	journal = {ACM Transactions on Computer-Human Interaction},
	author = {Draxler, Fiona and Werner, Anna and Lehmann, Florian and Hoppe, Matthias and Schmidt, Albrecht and Buschek, Daniel and Welsch, Robin},
	month = apr,
	year = {2024},
	pages = {1--40},
}

@inproceedings{10.1145/3706598.3713522,
author = {He, Jessica and Houde, Stephanie and Weisz, Justin D.},
title = {Which Contributions Deserve Credit? Perceptions of Attribution in Human-AI Co-Creation},
year = {2025},
isbn = {9798400713941},
publisher = {Association for Computing Machinery},
address = {New York, NY, USA},
url = {https://doi.org/10.1145/3706598.3713522},
doi = {10.1145/3706598.3713522},
booktitle = {Proceedings of the 2025 CHI Conference on Human Factors in Computing Systems},
articleno = {540},
numpages = {18},
location = {
},
series = {CHI '25}
}

@article{moruzzi_creative_2022,
	title = {Creative {Agents}: {Rethinking} {Agency} and {Creativity} in {Human} and {Artificial} {Systems}},
	volume = {9},
	issn = {2053-9320, 2053-9339},
	shorttitle = {Creative {Agents}},
	url = {https://www.tandfonline.com/doi/full/10.1080/20539320.2022.2150470},
	doi = {10.1080/20539320.2022.2150470},
	language = {en},
	number = {2},
	urldate = {2025-11-22},
	journal = {Journal of Aesthetics and Phenomenology},
	author = {Moruzzi, Caterina},
	month = jul,
	year = {2022},
	pages = {245--268},
}

@inproceedings{han_when_2024,
	address = {New York, NY, USA},
	series = {{CHI} '24},
	title = {When {Teams} {Embrace} {AI}: {Human} {Collaboration} {Strategies} in {Generative} {Prompting} in a {Creative} {Design} {Task}},
	copyright = {All rights reserved},
	isbn = {979-8-4007-0330-0},
	shorttitle = {When {Teams} {Embrace} {AI}},
	url = {https://dl.acm.org/doi/10.1145/3613904.3642133},
	doi = {10.1145/3613904.3642133},
	urldate = {2024-05-13},
	booktitle = {Proceedings of the {CHI} {Conference} on {Human} {Factors} in {Computing} {Systems}},
	publisher = {Association for Computing Machinery},
	author = {Han, Yuanning and Qiu, Ziyi and Cheng, Jiale and LC, RAY},
	month = may,
	year = {2024},
	pages = {1--14},
}

@inproceedings{chen_once_2025,
	address = {Cham},
	title = {Once {More} with (the {Right}) {Feeling}: {How} {Historical} {Fiction} {Writing} {Processes} of~{Character} {Design}, {Plot} {Outline}, and~{Context} {Checking} {Are} {Affected} by~{Co}-{Writing} with~{ChatGPT}},
	copyright = {All rights reserved},
	isbn = {978-3-031-92823-9},
	shorttitle = {Once {More} with~(the {Right}) {Feeling}},
	doi = {10.1007/978-3-031-92823-9_7},
	language = {en},
	booktitle = {{HCI} in {Business}, {Government} and {Organizations}},
	publisher = {Springer Nature Switzerland},
	author = {Chen, Yun and Wang, Yiwei and Chan, Antoni B. and Li, Jixing and LC, Ray},
	editor = {Siau, Keng Leng and Nah, Fiona Fui-Hoon},
	year = {2025},
	pages = {79--101},
}
